\documentclass[journal]{IEEEtran}

\usepackage{amsmath}
\usepackage{multirow}
\usepackage[caption=false,font=footnotesize]{subfig}
\usepackage{graphicx}
\usepackage{caption}
\usepackage{colortbl}
\usepackage{xcolor}
\usepackage{cite}
\usepackage[hyphens]{url}
\usepackage{amsfonts}
\usepackage{graphicx}
\usepackage{balance}

\newcommand\ourmethod{\textit{Accoustate}}

\newtheorem{definition}{Definition}
\begin{document}

\title{\textit{Accoustate}: Auto-annotation of IMU-generated Activity Signatures under Smart Infrastructure}

\author{Soumyajit Chatterjee, Arun Singh,
	Bivas Mitra,
	and Sandip Chakraborty,% <-this % stops a space
	\thanks{This work has been submitted to the IEEE for possible publication. Copyright may be transferred without notice, after which this version may no longer be accessible.}
	\thanks{Soumyajit Chatterjee, Bivas Mitra and Sandip Chakraborty are with the Department
		of Computer Science and Engineering, Indian Institute of Technology Kharagpur, West Bengal - 721302, India e-mail: (see soumyachat@iitkgp.ac.in).}
	\thanks{Arun Singh was with the Department of Computer Science and Engineering, Indian Institute of Technology Kharagpur, West Bengal - 721302, India.}}

\markboth{Journal of \LaTeX\ Class Files,~Vol.~14, No.~8, August~2015}%
{Chatterjee \MakeLowercase{\textit{et al.}}: Bare Demo of IEEEtran.cls for IEEE Journals}	

\maketitle

\begin{abstract}
Human activities within smart infrastructures generate a vast amount of IMU data from the wearables worn by individuals. Many existing studies rely on such sensory data for human activity recognition (HAR); however, one of the major bottlenecks is their reliance on pre-annotated or labeled data. Manual human-driven annotations are neither scalable nor efficient, whereas existing auto-annotation techniques heavily depend on video signatures. Still, video-based auto-annotation needs high computation resources and has privacy concerns when the data from a personal space, like a smart-home, is transferred to the cloud. This paper exploits the acoustic signatures generated from human activities to label the wearables' IMU data at the edge, thus mitigating resource requirement and data privacy concerns. We utilize acoustic-based pre-trained HAR models for cross-modal labeling of the IMU data even when two individuals perform simultaneous but different activities under the same environmental context. We observe that non-overlapping acoustic gaps exist with a high probability during the simultaneous activities performed by two individuals in the environment's acoustic context, which helps us resolve the overlapping activity signatures to label them individually. A principled evaluation of the proposed approach on two real-life in-house datasets further augmented to create a dual occupant setup, shows that the framework can correctly annotate a significant volume of unlabeled IMU data from both individuals with an accuracy of $\mathbf{82.59\%}$ ($\mathbf{\pm 17.94\%}$) and $\mathbf{98.32\%}$ ($\mathbf{\pm 3.68\%}$), respectively, for a workshop and a kitchen environment.
\end{abstract}

%\begin{CCSXML}
%	<ccs2012>
%	<concept>
%	<concept_id>10003120.10003138.10003142</concept_id>
%	<concept_desc>Human-centered computing~Ubiquitous and mobile computing design and evaluation methods</concept_desc>
%	<concept_significance>500</concept_significance>
%	</concept>
%	<concept>
%	<concept_id>10003120.10003138.10003141</concept_id>
%	<concept_desc>Human-centered computing~Ubiquitous and mobile devices</concept_desc>
%	<concept_significance>300</concept_significance>
%	</concept>
%	<concept>
%	<concept_id>10003120.10003138.10003141.10010895</concept_id>
%	<concept_desc>Human-centered computing~Smartphones</concept_desc>
%	<concept_significance>100</concept_significance>
%	</concept>
%	</ccs2012>
%\end{CCSXML}
%
%%\ccsdesc[500]{Human-centered computing~Ubiquitous and mobile computing design and evaluation methods}
%\ccsdesc[500]{Human-centered computing~Ubiquitous and mobile devices}
%%\ccsdesc[100]{Human-centered computing~Smartphones}
%


\section{Introduction}
Applications on Human Activity Recognition (HAR) are essential for developing any smart infrastructure, be it a smart home, a smart workplace, or a smart factory. There have been various endeavors on HAR~\cite{icpr_1} using time-series data captured from sensors like Inertial Motion Units (IMU) attached with individuals in different forms like a smartwatch or a smart band. However, these models primarily rely on supervised training that needs a huge volume of labeled data~\cite{thms}. Traditional approaches for data annotation of IMU streams use human-in-the-loop that not only produces noisy and unreliable labels~\cite{conflict} but is also a significant costly \& time-consuming process~\cite{buildsys20}, which does not even scale. In order to reduce human participation during data annotation, approaches like \textit{Active Learning}~\cite{active_main} and \textit{Experience Sampling} (ESM)~\cite{esm_main} have been adopted in several recent researches. However, these methods heavily depend on the choice of annotators~\cite{nroy1} and need a partially-labeled dataset for bootstrapping. 
Moreover, as human activities are a continuous time-series process, it is necessary to correctly identify the IMU boundaries where the subject changes her activity \& annotate them accordingly. Any human-based annotation is likely to fail for such precise labeling of the IMU data. The problems escalate for scenarios like a smart-home designed for elderly assistance~\cite{buildsys20}, where the human-based annotation is not only challenging but also infeasible up to a certain extent. Hence, automating the data annotation of IMU streams is necessary for such scenarios.

%\subsection{Existing Works and Their Limitations}
Addressing these challenges, attempts have been made in the literature~\cite{auto_new_1,auto2,sensor_labeling3} to develop frameworks that can automatically annotate the IMU data streams for HAR relying on the auxiliary modality present in the environment. The choice of the auxiliary modality is crucial in these cases, and most of these frameworks use videos as the preferred option. Although video provides a rich source of granular information, this modality is privacy-intrusive, costly, and hardly conducive to infrastructures like smart homes. In this paper, we leverage the acoustic signals as auxiliary modality, which can be captured in the environment through virtual personal assistants (VPA) and Smart Speakers. Audio as the auxiliary modality provides us multiple advantages over video. (a) There exist sophisticated models, like~\cite{ubicoustics} built over rich publicly available datasets such as \texttt{YouTube-8M}~\cite{youtube} or \texttt{urbansound8k}~\cite{salamon2014dataset}, which can efficiently recognize human activities with high accuracy, even when multiple activities are performed simultaneously. (b) Importantly, audio processing is much lightweight compared to the complex video processing techniques employing large deep learning models~\cite{gowda2020smart}. This paves the way for on-device audio processing, conducted at the edge of smart infrastructure, preserving data privacy. In an initial work~\cite{laso}, we developed a platform called \textit{LASO} that can annotate the user activities by exploiting the acoustic signals generated from those activities. However, \textit{LASO} is limited to correctly annotate the data when there is only a single user in the environment. 

Additional challenges exist for annotating IMU data in the presence of more than one user while using audio as the auxiliary modality. First, when more than one users perform activities simultaneously under the same environment, the audio signals from respective activity sources get convoluted. Albeit audio-based HAR models~\cite{ubicoustics} return a set of activities with the corresponding confidence scores; however, it is challenging to separate the activity boundaries directly from such models. This challenge mainly stems from external non-human noises, which force the model to detect spurious activities and activities with low confidence. Side by side, IMU streams produce highly fluctuating and noisy signals, which make it challenging to extract the activity boundaries directly from the IMU. Finally, in the presence of more than one user performing multiple activities simultaneously, the acoustic signal alone cannot determine who is doing what, which is essential to annotate the IMU data stream captured from the individual users.

Owing to these challenges, we extend \textit{LASO} in this paper for supporting dual user activity annotation from acoustic signatures. Although a dual user setup does not generalize the activity annotation problem for a complete multi-user setup, acoustic-based activity annotation over a dual user setup is not straightforward and is the foremost step towards this generalization. Further, in a realistic environment, it is seldom the scenario when many users perform activities simultaneously within a common acoustic setup; the dual user case can itself cover a significant portion of the use-cases~\cite{alemdar2013aras}.

\subsection{Our Contributions}
In contrast to the existing works, the contributions of this paper are as follow.\\

\noindent\textbf{(1) Unsupervised activity to user mapping in a dual user setup:} \ourmethod{} exploits the activity acoustic patterns in a dual user setup to map an activity to a user through unsupervised joint analysis of acoustic and IMU signals (Section~\ref{method}). The crux behind the design of the proposed method, called \ourmethod{}, is that the sounds produced from human activities are not continuous~\cite{chipaper} and are generated from specific granular micro-activities. For example, in a smart kitchen during the activity \textit{cooking}, no sound may be produced when the user is carrying a frying pan. Indeed, when the frying pan is used to fry something, a distinctive sound will be produced.\\

\noindent\textbf{(2) Unsupervised labeling of IMU data from pretrained acoustic-based HAR model:} \ourmethod{} intelligently utilizes the unsupervised nearest-neighbor model to annotate the activities associated with individual IMU change-points. For this purpose, we utilize pre-trained audio-based HAR models, which are lightweight and thus run over an edge device, preserving data privacy.\\

\noindent\textbf{(3) Implementation, deployment, and testing of \ourmethod:} We have implemented and tested \ourmethod{} over two different datasets, a Workshop and another Kitchen environments, and benchmarked the resource consumption profile of the running module to show its feasibility for deploying over edge devices under the smart infrastructure. The results show that \ourmethod{} generates annotated IMU data with an appreciable accuracy of $82.59\%$ ($\pm 17.94\%$) and $98.32\%$ ($\pm 3.68\%$), respectively for the two environments, for the cases when a correct activity to user mapping could be done ($10$ out of $12$ cases in Workshop, and $3$ out of $5$ cases in Kitchen) in a two-user simultaneous activity setup (Section~\ref{evaluation}).

\section{Related Work}
\label{relwork}
\begin{table*}[]
	\centering
	\scriptsize
	\caption{Summary of Related Work on Automatic Annotation of Sensor Modalities}
	\label{tbl:rel_work}	
	\begin{tabular}{|l|c|c|c|l|c|c|}
		\hline
		\multicolumn{1}{|c|}{\textbf{Paper}} &
		\textbf{\begin{tabular}[c]{@{}c@{}}Primary\\ Modalities\end{tabular}} &
		\textbf{Auxiliary Modalities} &
		\textbf{\begin{tabular}[c]{@{}c@{}}External\\ Labeling\\ Source\end{tabular}} &
		\multicolumn{1}{c|}{\textbf{Brief Methodology}} &
		\textbf{\begin{tabular}[c]{@{}c@{}}Label\\ General-Purpose \\ Activities?\end{tabular}} &
		\textbf{Multi-User?} \\ \hline
		Alirezaie et. al.~\cite{auto1} &
		\begin{tabular}[c]{@{}c@{}}Medical\\ Monitoring \\ Sensors\end{tabular} &
		NA &
		\begin{tabular}[c]{@{}c@{}}Open-Linked\\ Knowledge\\ Sources\end{tabular} &
		1. Analysis through \emph{Abductive} reasoning. &
		No &
		NA \\ \hline
		Benndorf et. al.~\cite{auto2} &
		\begin{tabular}[c]{@{}c@{}}IMU\\ Sensors\end{tabular} &
		Video &
		OpenPose~\cite{openpose} &
		\begin{tabular}[c]{@{}l@{}}1. Key-points extraction.\\ 2. Annotate sensors from videos.\end{tabular} &
		Basic Physical Activities &
		Yes \\ \hline
		Rey et. al.~\cite{sensor_labeling3} &
		\begin{tabular}[c]{@{}c@{}}IMU\\ Sensors\end{tabular} &
		\begin{tabular}[c]{@{}c@{}}Monocular\\ RGB videos\end{tabular} &
		YouTube Videos &
		\begin{tabular}[c]{@{}l@{}}1. Extracting 2D poses from Video Frames\\ 2. Based on a regression model\end{tabular} &
		Fitness Exercises &
		Yes \\ \hline
		RecycleML~\cite{crossmodallearning} &
		\begin{tabular}[c]{@{}c@{}}IMU\\ Sensors\end{tabular} &
		Audio/Video &
		NA &
		1. Cross-modal knowledge transfer &
		Basic Physical Activities &
		No \\ \hline
		LASO~\cite{laso} &
		\begin{tabular}[c]{@{}c@{}}IMU\\ Sensors\end{tabular} &
		Audio &
		YouTube-8M~\cite{youtube} &
		\begin{tabular}[c]{@{}l@{}}1. Cross-modal change detection\\ 2. Audio-based activity recognition.\end{tabular} &
		Yes &
		No \\ \hline
		\textit{\textbf{Accoustate}} &
		\begin{tabular}[c]{@{}c@{}}\textbf{IMU}\\ \textbf{Sensors}\end{tabular} &
		\textbf{Audio} &
		YouTube-8M~\cite{youtube} &
		\begin{tabular}[c]{@{}l@{}}1. Exploiting gaps in human activities\\ 2. Utilizing acoustic gaps\\ 3. Unsupervised mapping of activities\end{tabular} &
		\textbf{Yes} &
		\textbf{\begin{tabular}[c]{@{}c@{}}Two\\ Occupants\end{tabular}} \\ \hline
	\end{tabular}
\end{table*}
Obtaining annotations for the collected data has been a prime challenge in the field of HAR. Typical locomotive and inertial sensors generate millions of instances in a day. Subsequently, a majority of the data generated by smart infrastructures become unusable because of the unavailability of proper annotation~\cite{buildsys20,nroy1}. Also, given such a huge volume of data generated by these smart infrastructures, the standard approach of human annotation becomes infeasible (both in terms of cost and time). To counter these challenges several approaches like \textit{Active Learning}~\cite{active_main} and \textit{Experience Sampling}~\cite{esm_main} have been adopted. These approaches reduce the overall data to be annotated by polling the user only when the system is not confident about the generated label. However, a major drawback of these approaches is their heavy dependency on the human-in-the-loop based approach that requires a proper choice of annotators~\cite{nroy1} and can even be noisy in certain cases~\cite{conflict}. Furthermore, approaches like \textit{Active Learning} need a small set of pre-labeled data to start with, and obtaining such datasets for complex ADL(s) can be challenging~\cite{active1}. Similarly, for ESM-based approaches~\cite{esm}, the participants themselves may need to provide labels in situ. However, such a setup may not always be possible, especially in smart homes designed for assisting the elderly population.

Understanding these challenges, a different approach that has been adapted in recent times is the development of automated annotation frameworks~\cite{auto2,auto_new_1,auto_new_2,sensor_labeling3}. Notably, most of these works use one or more auxiliary modalities that assist the framework in generating the labels for the unlabeled primary sensor modality. In most of the cases, a typical choice for many of these approaches has been to use videos~\cite{auto2,sensor_labeling3} as an auxiliary modality, which provides very granular information regarding the basic physical activities (like walking, standing) and for specialized activities like fitness exercises. Additionally, the video also provides rich information to perform cross-modal information associations~\cite{idiot}, albeit it is highly privacy-invasive and computationally expensive to process. Understanding these challenges, recent works like~\cite{ubicoustics,electricalsensing} have pointed out different alternative modalities that can allow granular activity recognition in smart homes. Out of these, the acoustic context, in particular, can be used in a plug-and-play setup with diverse pre-trained models without using any external hardware~\cite{ubicoustics} or human intervention.

Interestingly, existing works like~\cite{crossmodallearning,laso} have used this idea for generating training data. However, these frameworks are designed explicitly for single-user scenarios and involve cross-modal detection of activities that may be difficult to map when multiple unlabeled IMU streams arrive in the system from more than one user along with a globally confounded audio stream. Additionally, the performance of frameworks like \textit{RecycleML}~\cite{crossmodallearning} heavily depends on the availability of small yet significant bootstrapping data, obtaining which can be challenging in real-life scenarios. The summary of comparison with the existing related works is shown in \tablename~\ref{tbl:rel_work}.
\section{Dataset}
\label{datasets}
%This section starts with defining the problem formally followed by the key idea behind our design and then discusses the dataset used to develop the model. 
%This section formally defines the problem statement and the key idea behind the development of \ourmethod{}. It then describes the entire data collection and data augmentation setup that we utilize to design and evaluate \ourmethod{}.
%We further assume that the label-space of the pre-trained audio-based HAR model includes the primary activities $p_i$ and $p_j$. 
%\begin{figure}[!ht]
%	\captionsetup[subfigure]{}
%	\begin{center}
%		\subfloat[Workshop\label{fig:working_workshop}]{
%			\includegraphics[width=0.48\linewidth,keepaspectratio]{figures/workshop_missing_annotations.png}
%		}
%		\subfloat[Kitchen\label{fig:working_kitchen}]{
%			\includegraphics[width=0.48\linewidth,keepaspectratio]{figures/kitchen_missing_annotations.png}
%		}
%	\end{center}
%	\caption{Percentage of IMU data with auxiliary activities}
%	\label{fig:auxiliary}
%\end{figure}
%\subsection{Challenges with the Open Datasets}
One of the major challenges of designing and testing \ourmethod{} is the unavailability of sufficient data from a realistic smart infrastructure environment. Data from commercial solutions such as Samsung Connected Living are not publicly available. On the contrary, publicly available datasets typically collect data over a very controlled and time-constrained environment. For example, the CMU-MMAC dataset\footnote{\url{http://kitchen.cs.cmu.edu/} (Accessed: \today)} contains activity data from a smart kitchen environment; however, the primary focus has been on individual body-parts' movement patterns while performing various micro-activities (for example, \textit{beating eggs}, \textit{taking fork}, etc. while \textit{cooking}).  Further, participants are connected with multiple sensors at different body parts, hindering them from free movements. Because of such constraints, the natural \& organic activity sequences, like bringing the frying pan from a distanced shelf to the oven, get hampered. However, such activity sequences play a crucial role while designing \ourmethod{}. Hence, we opt to rely on an in-house lab-scale smart environment, where individuals can perform the given tasks without external control and with minimum interference from the connected devices. 
%For this, we collect the single-user datasets and augment them by time-synchronizing the IMU signals and convoluting the acoustic signals among volunteers to create a synthetic dataset that can effectively mimic the multi-user setup with two users performing simultaneous activities. 
%The details of the setup follow.

\subsection{Data Collection in a Single User Setup}
We rely on a minimal setup where the IMU data is collected using a Moto-360 smartwatch (sampling rate = $50$Hz) worn on the preferred arm of the participant. A COTS Smartphone is utilized to capture the audio generated from the environment (sampling rate = $44.1$kHz). The smartwatch was paired apriori with the smartphone over Bluetooth for inter-modality synchronization. We obtain the data from two different environments, namely \textbf{Workshop} and \textbf{Kitchen}, involving a total of $8$ volunteers in this experiment. We collected the data in a single-user setup, where every volunteer has been asked to perform one single activity at a time, independently \& freely, without having any external constraints. In the \textbf{Workshop} environment, we involve $4$ volunteers and ask them to separately perform two \emph{primary} activities -- (a) \textit{hammering} a wooden plank or a metal pipe, \& (b) cutting a wooden plank or metallic pipe \textit{using a saw}. During this process, the participants organically perform other auxiliary micro-activities, like picking up the nails, or fitting the plank, etc. Similarly in \textbf{Kitchen} environment, the other $4$ volunteers are asked to separately conduct two \emph{primary} activities -- (a) \textit{chopping} vegetables with a knife on a chopping board, \& (b) \textit{cooking} the vegetables in a frying pan or a cast iron wok. For both the \textbf{Workshop} and the \textbf{Kitchen} environments, we captured timestamped videos with a frame rate of $30$fps and used them to annotate the ground-truth activity labels. 
%This is important to note that, even if we had not explicitly instructed the volunteers to take breaks during the tasks, \figurename~\ref{fig:auxiliary} depicts that volunteers switch from primary to auxiliary activity. 

It can be noted that this data has been collected during the design of \textit{LASO}~\cite{laso}. Unfortunately, due to the COVID-19 pandemic, we could not extend these experiments in a multi-user setup. So, this paper applied a judicial data augmentation mechanism using a signal convolution technique to synthesize the multi-user data from the available single-user activity data. 
%We ensured that this data augmentation does not collude with the experimental results by oversimplifying the activity patterns. The detail follows.

\begin{table}[]
	\scriptsize
	\centering
	\caption{Augmented Dataset Details -- \textit{Workshop}}
	\label{table:workshop_dataset}
	\begin{tabular}{|c|c|c|c|c|}
		\hline
		\multirow{2}{*}{\textbf{Combination Id}} &
		\multicolumn{2}{c|}{\textbf{Activities}} &
		\multirow{2}{*}{\textbf{IMU Instances}} &
		\multirow{2}{*}{\textbf{\begin{tabular}[c]{@{}c@{}}Global Audio\\ Duration (secs)\end{tabular}}} \\ \cline{2-3}
		& \textbf{Hammer} & \textbf{Saw} &      &         \\ \hline
		C1  & U1              & U3           & 6730 & 134.739 \\ \hline
		C2  & U1              & U2           & 2629 & 52.628  \\ \hline
		C3  & U1              & U4           & 7652 & 153.222 \\ \hline
		C4  & U3              & U1           & 6957 & 139.27  \\ \hline
		C5  & U4              & U1           & 6957 & 139.27  \\ \hline
		C6  & U2              & U1           & 6957 & 139.27  \\ \hline
		C7  & U4              & U3           & 6730 & 134.739 \\ \hline
		C8  & U4              & U2           & 2629 & 52.628  \\ \hline
		C9  & U2              & U3           & 6730 & 134.739 \\ \hline
		C10 & U3              & U2           & 2629 & 52.628  \\ \hline
		C11 & U3              & U4           & 7652 & 153.222 \\ \hline
		C12 & U2       
		
		& U4           & 7652 & 153.222 \\ \hline
	\end{tabular}
\end{table}

\begin{table}[]
	\scriptsize
	\centering
	\caption{Augmented Dataset Details -- \textit{Kitchen}}
	\label{table:kitchen_dataset}
	\begin{tabular}{|c|c|c|c|c|}
	\hline
	\multicolumn{1}{|l|}{\multirow{2}{*}{\textbf{Combination Id}}} &
	\multicolumn{2}{c|}{\textbf{Activities}} &
	\multirow{2}{*}{\textbf{IMU Instances}} &
	\multirow{2}{*}{\textbf{\begin{tabular}[c]{@{}c@{}}Global Audio\\ Duration (secs)\end{tabular}}} \\ \cline{2-3}
	\multicolumn{1}{|l|}{} & \textbf{Cooking} & \textbf{Chopping} &      &         \\ \hline
		C1 & U3               & U1                & 20983 & 420.305 \\ \hline
		C2 & U2               & U1                & 24351 & 487.787 \\ \hline
		C3 & U3               & U2                & 20981 & 420.305 \\ \hline
		C4 & U4               & U2                & 11596 & 232.304 \\ \hline
		C5 & U4               & U1                & 11597 & 232.304 \\ \hline
	\end{tabular}
\end{table}

\subsection{Realistic Data Augmentation for a Multi-user Setup}
We use the collected single-user datasets and augment them by time-synchronizing the IMU signals and convoluting the acoustic signals among volunteers to create a synthetic dataset that can effectively mimic the multi-user setup with two users performing simultaneous activities. Before creating the augmented dataset, we first noise profile each audio file for noise reduction, albeit we keep the IMU data unfiltered. Next, we use Audacity~\cite{audacity} to convolute the time-series acoustic signals from two different volunteers performing two different activities, ensuring that the audio processing steps maintain the overall quality and standard of the output files. Next, for the IMU data and the ground-truth activity timings, we synchronize them considering a global time reference by choosing the earliest time among the two volunteers in the combination. As the IMU sampling rate is fixed (at $50$Hz) and the IMU data for both the volunteers are synchronized, the final augmented dataset will have an approximately equal number of IMU instances for each volunteer. In this context, it is important to note that this augmentation strategy only creates a virtual setup with two occupants and does not tamper with the existing nature of the two real-life datasets. Following this data augmentation strategy, we create a set of $12$ (\tablename~\ref{table:workshop_dataset}) and $5$ (\tablename~\ref{table:kitchen_dataset}) unique virtual combinations for each subject-activity pair in the two setups, respectively.
\section{Solution Overview}
\label{method}
\begin{figure}[!t]
	\includegraphics[width=0.86\linewidth, keepaspectratio]{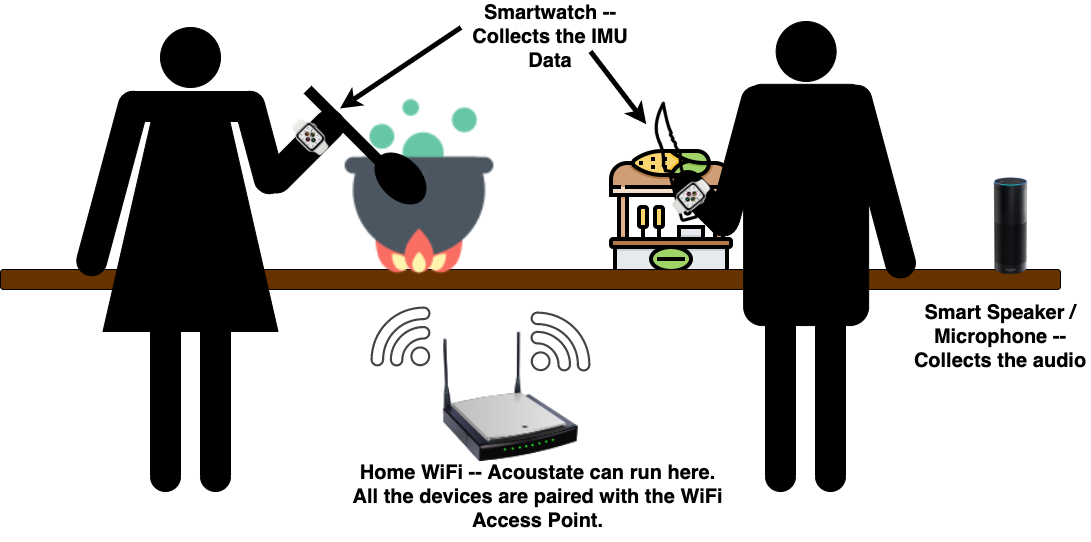}
	\caption{\ourmethod{} Working Environment}
	\label{fig:acoustatew}
\end{figure}

%This section starts with defining the problem formally. 
%Focusing further on these stops and analyzing the changes in the input modalities, we observe that these opportunistic stops indeed create significant changes in the observed signatures. From \figurename~\ref{fig:audio_change}, we observe that whenever a user stops, the signal changes significantly with occasional changes due to external noise. A similar analysis on the frequency domain characteristics of the augmented acoustic signatures (See \figurename~\ref{fig:mixed_audio_change}) shows the frequency signatures change significantly once one activity stops, albeit the presence of external noise and presence of another source complicates the detection.
%Concerning the IMU signatures (See \figurename~\ref{fig:imu_change_pilot}), the analysis reveals a similar behavior, albeit with a slightly smaller window compared to the acoustic change. This is because when the user stops performing the activity, the acoustic signature drops immediately, while the IMU signatures still record the transition. For example, when a user resumes using a saw, the acoustic signature captures this instantly; however, the change starts a bit earlier for the IMU signature as the user may take the saw and prepare for resuming the activity.
%\subsection{Problem Definition}
Let two users $\mathcal{U}_m$ and $\mathcal{U}_n$ perform two different primary activities $p_i$ (denoted as $\mathcal{U}_m \rightarrow p_i$) and $p_j$ (denoted as $\mathcal{U}_n \rightarrow p_j$, respectively, for a time period $[0,\mathcal{T}]$. Let $\mathcal{I}_m(0, \mathcal{T})$ and $\mathcal{I}_n(0, \mathcal{T})$ be the unlabeled IMU data collected through the wearable from each of the two users (see \figurename~\ref{fig:acoustatew}), respectively, and $\mathcal{A}(0, \mathcal{T})$ be the audio signal for the entire duration $[0, \mathcal{T}]$ captured from a VPA deployed in the environment. 
%We formally define the problem as follows. 

\subsection{Problem Definition and Key Idea}
The objective of \ourmethod{} is to develop a framework to annotate the completely unlabeled IMU data $\mathcal{I}_n(0, \mathcal{T})$ and $\mathcal{I}_m(0, \mathcal{T})$ utilizing the acoustic information extracted from $\mathcal{A}(0, \mathcal{T})$. Our key idea is that there are granular auxiliary micro-activities, say $\{a^1_m, a^2_m, \hdots\}$, within a primary activity $p_i$, some of which does not generate a distinctive sound, thus producing a gap in the acoustic signal. Accordingly, We define a term \textit{``Acoustic Gaps''} as follows.
%We consider that a user $\mathcal{U}_m$ may take intermediate breaks during her primary activity $p_i$ and perform different other activities, say $\{a^1_m, a^2_m, \hdots\}$, called the \textit{auxiliary activities} (say \textit{``taking a rest''} or \textit{``drinking water''} while conducting a primary activity \textit{``cooking''}). Let the duration of an auxiliary activity being performed by $\mathcal{U}_m$ be $[t, t+\Delta]$, where $0 \le t < t+\Delta \le \mathcal{T}$. \ourmethod{} considers that primary activities generate distinctive sounds, whereas auxiliary activities may (ex. \textit{``tapping water''}) or may not (ex. \textit{``drinking water''}) produce a sound. We define a term \textit{``Acoustic Gaps''} as follows.
\begin{definition}[Acoustic Gap]\label{def:stop_duration}
	Let a user $\mathcal{U}_m$ performs a primary activity $p_i$ for a period $[0,\mathcal{T}]$. We define an acoustic gap as the intermittent time duration $[t, t+\Delta]$, ($\Delta \ge 1$s), when a different acoustic context is produced due to an interleaved auxiliary micro-activity $a^k_m$. 
    %when the user takes a break from $p_i$ and perform an auxiliary activity $a^k_m$, thus producing a different acoustic context intermittently.
\end{definition}
We aim to leverage the acoustic gaps extracted from $\mathcal{A}(0, \mathcal{T})$ during primary activities $\{p_i, p_j\}$ to opportunistically annotate the completely unlabeled IMU data $\mathcal{I}_n(0, \mathcal{T})$ and $\mathcal{I}_m(0, \mathcal{T})$. We consider that the label-space of the primary activities $\{p_i, p_j\}$ is known and included in the label-space of a pre-trained audio-based HAR model like~\cite{ubicoustics}. However, the user-to-activity mappings $\mathcal{U}_m \rightarrow p_i$ and $\mathcal{U}_n \rightarrow p_j$ are not known. Similar to previous works like~\cite{crossmodallearning}, we also assume that IMU and audio signals are time-synchronized using any standard approaches like RTP or NTC~\cite{cmummac,grenoble}. 

\begin{figure}[!t]
	\includegraphics[width=\columnwidth,keepaspectratio]{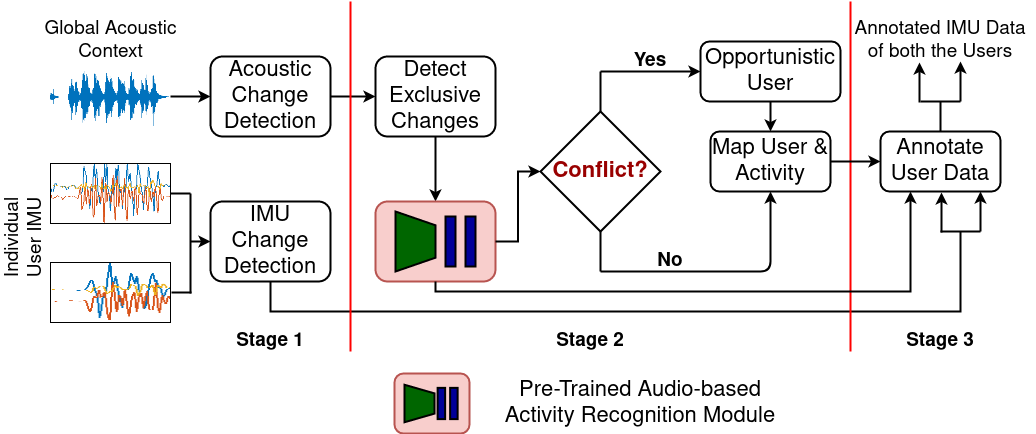}
	\caption{Framework Overview}
	\label{fig:framework}
\end{figure}
%Assume there are two users $\mathcal{U}_m$ and $\mathcal{U}_n$ who are performing simultaneous but different primary activities under the same smart infrastructure. We consider that within a time interval $[0,\mathcal{T}]$, each user $\mathcal{U}_m$ performs a single primary activity $p_m$ ( say, \textit{cooking}); however, $\mathcal{U}_m$ may take a break while doing the primary activity, when it perform some auxiliary activities $\{a^1_m, a^2_m, \hdots\}$ (say, \textit{drinking water} or \textit{resting the hand} while \textit{cooking}). The auxiliary activities may or may not produce any sound; however, we assume that the primary activities always produce some sound. Let $\mathcal{I}_m(0, \mathcal{T})$ and $\mathcal{I}_n(0, \mathcal{T})$ be the IMU signals from $\mathcal{U}_m$ and $\mathcal{U}_n$, respectively. Further, let $\mathcal{A}(0, \mathcal{T})$ be the acoustic signal captured from the common environment under the smart infrastructure. 
\subsection{\ourmethod{} in a Nutshell}
By exploiting the acoustic gaps, \ourmethod{} first maps the primary activities $\{p_i, p_j\}$ to the users $\{\mathcal{U}_m, \mathcal{U}_n\}$ and then annotates the IMU data $\mathcal{I}_m(0, \mathcal{T})$ and $\mathcal{I}_n(0, \mathcal{T})$ with the corresponding primary activities $\{p_i \rightarrow \mathcal{U}_m, p_j \rightarrow \mathcal{U}_n\}$. As shown in \figurename~\ref{fig:framework}, the entire framework is divided into three major stages. In \textbf{Stage 1}, \ourmethod{} independently detects signal changes in $\mathcal{I}_{.}(0, \mathcal{T})$ and $\mathcal{A}(0, \mathcal{T})$ using unsupervised approaches. In \textbf{Stage 2}, \ourmethod{} extracts the acoustic gaps from the signal changepoints to map the primary activities to the corresponding users, i.e. $p_i \rightarrow \mathcal{U}_m$ and $p_j \rightarrow \mathcal{U}_n$. Specifically, \ourmethod{} relies on a pre-trained audio-based activity recognition module for identifying the primary activities $\{p_i, p_j\}$, which allows us to avoid human intervention and detect activities in an automated manner. However, lack of an appropriate number of acoustic gaps and presence of multiple acoustic sources may confuse \ourmethod, and thus the framework may generate conflicting mappings. To resolve this, \ourmethod{} applies \textit{a conflict resolution technique} that allows it to judiciously map an activity label to a user during the entire duration $[0, \mathcal{T}]$. Ultimately, in \textbf{Stage 3}, \ourmethod{} collates all the information from the previous two stages to finally output the annotated IMU data for both the users. The details of each of these steps follow.

\section{Stage 1: Signal Change Detection}
%The primary hypothesis behind the design of \ourmethod{} is that a user while performing an activity, takes momentary breaks once in a while, which produces acoustic gaps for that user. When two users perform different activities simultaneously under the same smart infrastructure (having the same environmental context), the acoustic gap from one user can be used to extract the acoustic signature of the other user to identify the corresponding activity label uniquely. Ideally, the IMU data from a user should also indicate an activity change when an acoustic gap is observed for that user (say, a change from ``\textit{cooking}'' to ``\textit{resting hand}''). Therefore, \ourmethod{} uses an unsupervised activity change detection over both the IMU and the acoustic signals; the details follow. 
The first stage of \ourmethod{} uses an unsupervised approach to detect the instances when the distributions of the input signals, both for $\mathcal{I}_{.}(0, \mathcal{T})$ and $\mathcal{A}(0, \mathcal{T})$, show a change in their patterns, indicating that an activity change has happened in the environment for one of the users. 
%As both the input modalities $\mathcal{I}_{.}(0, \mathcal{T})$ and $\mathcal{A}(0, \mathcal{T})$ do not contain any preassigned labels, we rely on a complete unsupervised approach to detect such pattern changes. 
%The detail follows. 

\subsection{Detecting Changes in IMU}
%The primary objective of this framework is to exploit stops to generate annotations for the unlabeled IMU data. 
%Since the IMU data is completely unlabeled, we must rely on some unsupervised approach to find when a user takes a break during an activity. One possible way to do this is to segment the entire IMU data from a user into windows, such that each window is based on the change of activities. 
%To detect the changes in the IMU data, \ourmethod{} segments $\mathcal{I}_{.}(0, \mathcal{T})$ for a user $\mathcal{U}_{.}$ for the duration $[0, \mathcal{T}]$ when it performed a primary activity under a smart infrastructure. 
The objective of this step is to find out windows of duration $[\nu, \eta], 0 \leq \nu < \eta \leq \mathcal{T}$ for each user $\mathcal{U}_u$, such that each $\mathcal{I}_u(\nu, \eta)$ corresponds to one of the activities from $\{p_u, a^1_u, a^2_u, \hdots\}$. 

\subsubsection{Calculating Change-Points}
To achieve this, we first rely on the statistical \textit{Change-point Detection}~\cite{change_point2,hankel} approach to evaluate the changes in the IMU data stream. Formally, a change-point represents a point in time where a time series or a stochastic process has changed its probability distribution. Change-point scores quantify these changes in terms of numerical scores. \ourmethod{} computes the change-point scores to quantify the change in the IMU data stream considering the input from a tri-axial accelerometer. Say, $x_t$ represents the input from a tri-axial accelerometer at time $t$. Then, to compute the change-point scores, we create windows in the IMU data, such that a window $X_t = \{x_t, x_{t+1}, \hdots, x_{t+f-1}\}$, where, $f$ is the window size~\footnote{In this paper, we empirically set the value of $f = 25$ (time window of $0.5$ secs) for computing the change-point scores}. Subsequently, we compute the change-point score $\mu_t$ between any two consecutive IMU windows $X_t$ and $X_{t+f}$ using the $\alpha$-relative Pearson Divergence Estimation (PE), following a similar procedure as discussed in~\cite{smartsegment}.

%Let, $\mathcal{P}$ and $\mathcal{P}'$ be the two statistical distribution. Then, the absolute values of $\alpha$-PE divergence between $\mathcal{P}$ and $\mathcal{P}'$, denoted as $\delta(\mathcal{P}|\mathcal{P}')$ is computed as follows.  
%\begin{equation}
%\label{delta}
%\delta(\mathcal{P}|\mathcal{P}') = \frac{1}{2}\int p'_\alpha(X) \left(\frac{p(X)}{p'(X)}-1\right)^2 \times dX,\: \alpha \in (0,1)
%\end{equation} 
%where, $p(X)$ and $p'(X)$ are the probability densities of the two distributions $\mathcal{P}$ and $\mathcal{P}'$, respectively, and $p'_{\alpha}(X) = \alpha{p(X)} + (1-\alpha){p'(X)}$ is the mixture density. We use the standard relative density estimator (RuLSif) algorithm~\cite{rulsif} to numerically compute Eq.~(\ref{delta}). 
%
%Finally the \textit{change-point} score between the two windows $\mathcal{P}_{t}$ and $\mathcal{P}_{t+f}$, denoted as $\mu_t$, is obtained using the following expression.
%\begin{equation}
%\mu_t = \delta(\mathcal{P}_{t}|\mathcal{P}_{t+f}) + \delta(\mathcal{P}_{t+f}|\mathcal{P}_{t})
%\end{equation}

\subsubsection{Identifying the Actual Changes}
\begin{figure}[]
	\captionsetup[subfigure]{}
	\begin{center}
		\subfloat[\label{fig:imu_change_scores}]{
			\includegraphics[width=0.48\linewidth,keepaspectratio]{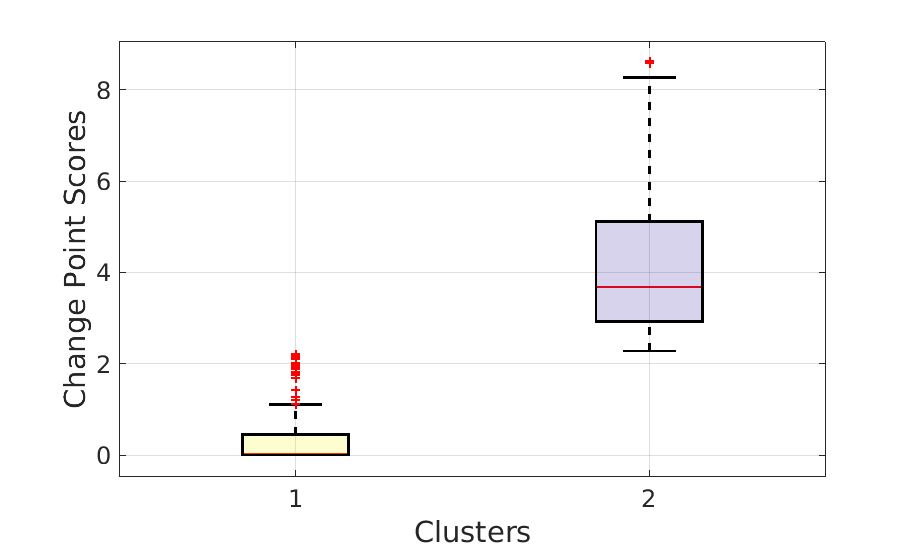}
		}
%		\subfloat[\label{fig:audio_cluster_index}]{
%			\includegraphics[width=0.33\linewidth,keepaspectratio]{figures/cluster_score.png}
%		}
		\subfloat[\label{fig:audio_cluster_scores}]{
			\includegraphics[width=0.48\linewidth,keepaspectratio]{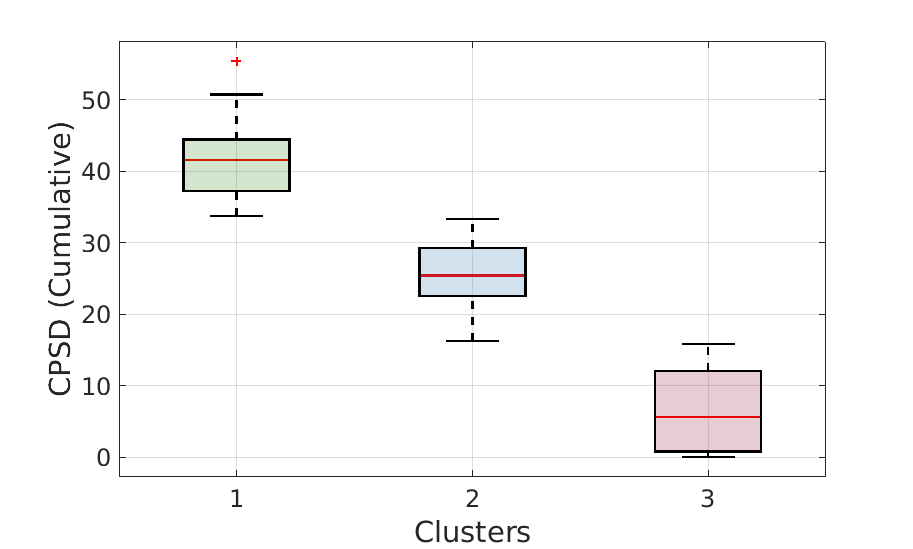}
		}
	\end{center}
	\caption{Analysis of change-points -- (a) Clustering the change-point scores for IMU data and (b) Clustering the CPSD values into an optimal number of clusters.}
\end{figure}

Although the \textit{change-point} scores can quantify the amount of changes in an IMU stream; however, they do not explicitly demarcate the actual event changes; instead, they give a relative score of changes in the distribution. Ideally, the actual activity changes within the IMU signal should produce relatively higher values of the change-point scores. However, as the IMU data is unlabeled, we cannot determine an empirical threshold on the change-point score. To solve this problem, we apply an unsupervised approach, where we cluster the obtained change-point scores into two sets -- one corresponding to the actual activity changes in the IMU data and the other containing the rest of the change-point scores. Thus, we perform a $k$-means clustering (with $k=2$, \figurename~\ref{fig:imu_change_scores}) on all the scores obtained from pairwise consecutive IMU windows. Subsequently, we demarcate the set of scores belonging to the cluster with a higher mean as the actual activity change scores.

%\todosoumya{Include a box plot for this in \figurename 5.}
%Notably, the score data may also contain some outliers; for example, a very sudden change in the hand movement can lead to a very high change-point score. It has been seen that more straightforward clustering approaches like $k$-means often tend to perform poorly in the presence of such outliers. We mitigate this problem by applying a standard outlier detection scheme, where the scores that deviate from the overall mean by more than two times the standard deviation are marked as outliers. However, one crucial point in this context is regarding the outliers present in the higher range of the distribution. As mentioned in the example, a sudden drastic change in the hand movement pattern can lead to a high change-point score. Since such changes can also point to the sudden breaks in the activity, we do not remove these outliers and instead demarcate these scores as changes for the remaining analysis.

\subsection{Detecting Changes in Acoustic Data}
We split the audio signal into $1$ second segments for extracting the audio change points and compare the two consecutive segments to detect changes. However, measuring the change in acoustic context is not straightforward, as the environmental acoustic context is a convoluted signal from different activities and other acoustic sources~\cite{soundadapter}. Therefore, existing methods like~\cite{crowdpp}, which use techniques like Mel Frequency Cepstral Coefficients (MFCC), fail to locate the changes correctly in our context. It is known that MFCC  becomes ineffective in the presence of multiple noise sources~\cite{mfcc_noise_2}, especially for non-speech environmental acoustic signatures~\cite{app6050143}. 

\subsubsection{Observing Changes in the Environmental Acoustic Context}
\begin{figure}[]
	\captionsetup[subfigure]{}
	\begin{center}
		%		\subfloat[\label{fig:audio_change}]{
		%			\includegraphics[width=0.33\linewidth,keepaspectratio]{figures/pilot_audio_1.png}
		%		}
		\subfloat[\label{fig:mixed_audio_change}]{
			\includegraphics[width=0.48\linewidth,keepaspectratio]{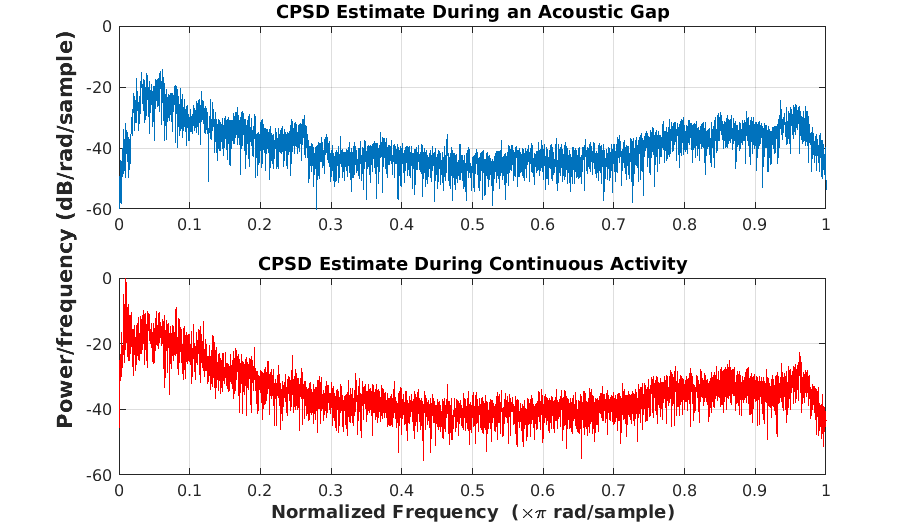}
		}
		\subfloat[\label{fig:imu_change_pilot}]{
			\includegraphics[width=0.48\linewidth,keepaspectratio]{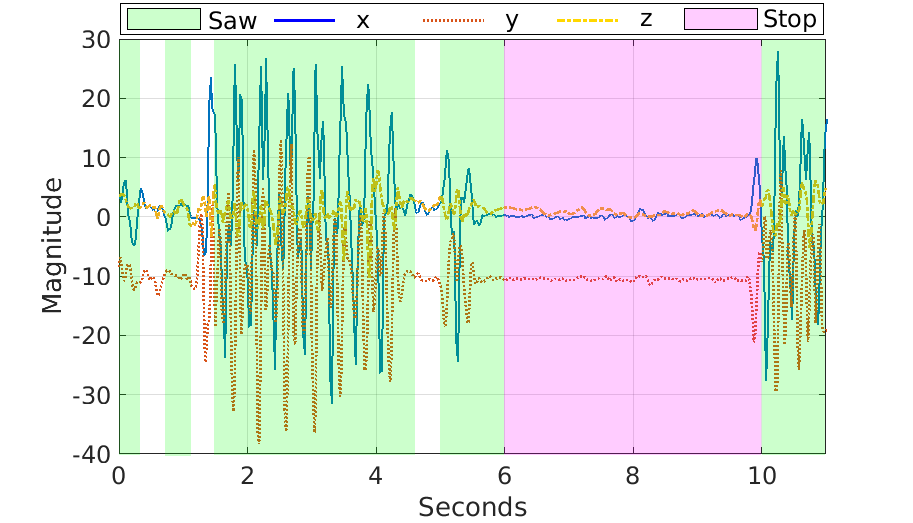}
		}
	\end{center}
	\caption{Observations during an acoustic gap -- (a) CPSD estimate and (b) Change in the IMU signatures}
\end{figure}

Notably, recent works like~\cite{app6050143} have pointed out that power spectral densities can appear as a more relevant alternative to cases where there are limitations on the usage of MFCC due to random noise components introduced by hardware or external sources. Based on these understandings, we choose~\emph{Cross Power Spectral Density} (CPSD) to identify changes in the environmental acoustic context. Formally, CPSD estimates the similarity (or relationship) between two time-domain signals~\cite{cpsd} and is obtained by performing Fourier transform on the cross-correlation of the two signals.
%Mathematically, CPSD $\psi_{uv}$, between two random time-domain signals $u$ and $v$, can be expressed as follows.
%\begin{equation}
%\psi_{uv} = \int_{-\infty}^{\infty}\rho_{uv}(\tau) e^{-j\omega\tau} d\tau
%\end{equation}
%where,
%\begin{equation}
%\rho_{uv}(\tau) = \lim_{T\rightarrow\infty}\frac{1}{T}\int_{0}^{T}u(t)v(t-\tau)dt
%\end{equation}
Since CPSD returns the power density across all the frequency bins present in both the signals, we consider the sum of absolute output values across all frequency bins to obtain the final CPSD value between two consecutive audio segments. In this context, an important observation is that unlike change-point scores computed for IMU signatures, higher CPSD values indicate lesser chances of any change between two consecutive segments. As \figurename~\ref{fig:mixed_audio_change} indicates, the CPSD values are marginally lower for the majority of the frequency values during an acoustic gap (when an activity change occurs) compared to the scenario when both the activities are being performed continuously. However, these are also mere values only, and we need to cluster them to demarcate actual changes. The detail follows.

\subsubsection{Identifying the Actual Changes in the Acoustic Context}
Similar to clustering the change-point scores for demarcating activity changes in the IMU data, here as well, we cluster the CPSD values. However, a primary problem with acoustic signatures is the presence of different random noise components. Additionally, we also observe that some acoustic signature changes are also introduced due to the behavioral changes in performing the activity. For example, a user may change how she holds the saw according to her ease in a workshop environment. Depending on that, the sound generated while cutting the plank can also change. Due to all these reasons, there can be a huge number of uneventful change-points if we cluster them into two sets only, which can negatively impact {\ourmethod}'s performance. Thus to avoid this, we first obtain the optimal number of clusters for CPSD values using the Silhouette score~\cite{rousseeuw1987silhouettes}. For this, we cluster the CPSD values into $c$ clusters, where $c \in [2,\mathcal{C}]$~\footnote{We choose $\mathcal{C}$ = $6$.} and choose $c$ with maximum Silhouette score as the optimal number of clusters. Once the clustering is done, we mark the cluster with minimum mean CPSD value as the cluster containing actual activity changes.

\section{Stage 2: Activity to User Mapping}
\ourmethod{} maps primary activities $\{p_i, p_j\}$ to users $\{\mathcal{U}_m, \mathcal{U}_n\}$ in two stages.  (a) Using  $\mathcal{I}_u, u \in \{m, n\}$, it first identifies $\mathcal{U}_u, u \in \{m, n\}$ who might have taken a break in her primary activity during the time segment $[\nu, \eta], 0 \leq \nu < \eta \leq \mathcal{T}$ (thus produces an acoustic gap in $\mathcal{A}(\nu, \eta)$ for $\mathcal{U}_u, u \in \{m, n\}$). (b) In the next step, it uses $\mathcal{A}(\nu, \eta)$ to obtain the activity labels from a pre-trained audio-based HAR model~\cite{ubicoustics} and maps those activities to the individual users based on the timing analysis over acoustic gaps. 
%The details follow. 

\subsection{Extracting Acoustic Gaps}
One of the major challenges in correlating IMU and Audio is that the change-points computed individually from them are not time-synchronized. \figurename~\ref{fig:imu_change_pilot} indicates that a difference is observed in the IMU signal whenever there is a change in the acoustic signal, albeit with a slightly smaller window compared to the acoustic change. This is because when the user stops performing the primary activity, the acoustic signature drops immediately, while the IMU signatures still record the transition. For example, when a user resumes using a saw, the acoustic signature captures this instantly; however, IMU changes a bit earlier when the user just picks up the saw. Therefore, \ourmethod{} uses an opportunistic approach to exploit the acoustic gaps by combining the observations from both modalities. For this, we use the notion of \textit{Exclusive Change} defined as follows. 
\begin{definition}[Exclusive Change]\label{def:exclusive_change}
	Say, at some time-interval $[\nu, \eta], 0 \leq \nu < \eta \leq \mathcal{T}$, we observe a change in $\mathcal{I}_m(\nu, \eta)$ for user $\mathcal{U}_m$. We define this change as an \textit{exclusive change} if and only if the following two conditions are met.
	\begin{enumerate}
		\item $\exists$ a time-interval $[\theta,\zeta], 0 \le \theta \le \nu < \eta \le \zeta \le \mathcal{T}$, where there is a change in $\mathcal{A}(\theta, \zeta)$. 
		\item For the entire time interval $[\theta,\zeta]$, we do not observe any change in $\mathcal{I}_n(\theta, \zeta)$ for the other user $\mathcal{U}_n$ present in the environment.
	\end{enumerate} 	
\end{definition}
\begin{figure}[!ht]
    \centering
    \includegraphics[width=0.70\linewidth]{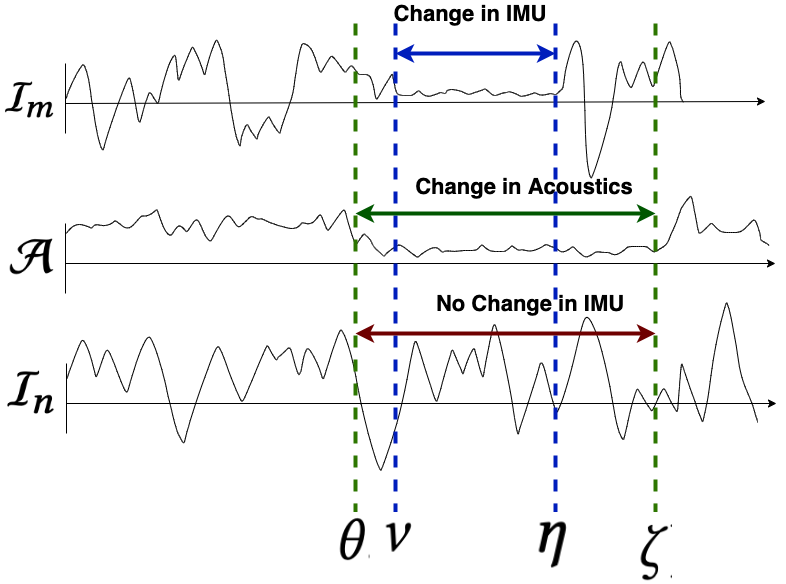}
    \caption{Exclusive Changes: $\mathcal{I}_m(\nu, \eta)$ indicates an exclusive change for the user $\mathcal{U}_m$}
    \label{fig:cp}
\end{figure}

Based on this definition, \figurename~\ref{fig:cp} shows an exclusive change for the user $\mathcal{U}_m$ over the IMU signal $\mathcal{I}_m$. The exclusive changes indicate the presences of an acoustic gap. Once we determine these exclusive changes for the individual users, we identify the activity labels from the acoustic context and map them to individual users as follows. 

\subsection{Mapping Activities to Users}
The acoustic gaps during the exclusive changes help us to find out a unique mapping from the activity labels $\{p_i, p_j\}$ to the users $\{\mathcal{U}_i, \mathcal{U}_j\}$. Let, $[\beta,\gamma]$ be a continuous time interval, and $[\nu, \eta], \beta \le \nu < \eta \le \gamma$ be an exclusive change detected for the user $\mathcal{U}_m$ from $\mathcal{I}_m(0, \mathcal{T})$. Let $\mathcal{A}_m(\beta, \gamma)$ and $\mathcal{A}_n(\beta, \gamma)$ be the pure acoustic signal components generated from the activities (primary or auxiliary) being performed by $\mathcal{U}_m$ and $\mathcal{U}_n$, respectively, during the time interval $[\beta,\gamma]$. Further, consider that $\mathcal{N}(\beta, \gamma)$ is the environmental noises generated from non-human activities (like the sound from AC, dog barking, etc.). Then, $\mathcal{A}(\beta, \gamma) = \mathcal{A}_m(\beta, \gamma) \oplus \mathcal{A}_n(\beta, \gamma) \oplus \mathcal{N}(\beta, \gamma)$, where $\oplus$ is the signal convolution operator. It can be noted that $\mathcal{A}(\beta, \gamma)$ should contain change-points near $\nu$ and $\eta$, as the primary activity for $\mathcal{U}_m$ changes at $\nu$ and $\eta$; therefore, $\mathcal{A}_m(\beta, \nu+\Delta_1)$, $\mathcal{A}_m(\nu+\Delta_1, \eta+\Delta_2)$, and $\mathcal{A}_m(\eta+\Delta_2, \gamma)$ should have different distributions in their \textit{Power Spectral Density} (PSD). Here, $\Delta_1$ and $\Delta_2$ are small adjustment windows, as the IMU change-points and the acoustic change-points may not be perfectly time-synchronized. However, $\mathcal{U}_n$ does not change her activity during $[\beta,\gamma]$, and therefore $\mathcal{A}_n(\beta, \gamma)$ should not ideally contain any change-points. 

\noindent\textbf{Activity detection from acoustic context:} We first detect the activities from the environmental acoustic signature $\mathcal{A}(\beta, \gamma)$ during the entire duration $[\beta,\gamma]$. As one of the main objectives of \ourmethod{} is to minimize human-in-the-loop, therefore, we rely on the concept of pre-trained models for audio-based activity recognition. Specifically, we adapt the model suggested in~\cite{ubicoustics} for this purpose, which is pre-trained using YouTube-8M~\cite{youtube} dataset and uses context information to filter out noisy activity labels effectively. For an input acoustic signature, the pre-trained model returns a set of detected activities with an associated confidence level indicating the model's confidence over the detected activities. Notably, the label space of~\cite{ubicoustics} also contains the workshop and kitchen activities defined in our datasets.

\noindent\textbf{Separating activities:} Let $\mathbb{A}(\beta, \gamma)$ be the set of activities returned by the pre-trained model~\cite{ubicoustics} during the time interval $[\beta,\gamma]$. $\mathbb{A}(\beta, \gamma) = \mathbb{A}_m(\beta, \gamma) \cup \{p_j\} \cup \mathbb{A}_{\mathcal{N}}(\beta, \gamma)$, where $\mathbb{A}_m(\beta, \gamma)$ is the set of activities (including the primary activity and the auxiliary activities) being performed by $\mathcal{U}_m$, $p_j$ is the single primary activity being performed by $\mathcal{U}_n$ during the entire duration $[\beta,\gamma]$, and $\mathbb{A}_{\mathcal{N}}(\beta, \gamma)$ is the set of non-human noisy activities. $\mathbb{A}_{\mathcal{N}}(\beta, \gamma)$ is easily separable as they do not belong to the target activity set. $\mathbb{A}_m(\beta, \gamma)$  should contain one primary activity $p_i$ within the duration $[\beta, \nu+\Delta_1]$ \& also within $[\eta+\Delta_2, \gamma]$, and an auxiliary activity $a_m$ within the duration $[\nu+\Delta_1, \eta+\Delta_2]$, as detected from the pre-trained acoustic-based activity recognition model~\cite{ubicoustics}. 

\noindent\textbf{Mapping the primary activities based on the IMU changes:} To map the primary activities $p_i$ and $p_j$ with the corresponding users $\mathcal{U}_m$ and $\mathcal{U}_n$, we now look into the change-points detected in $\mathcal{I}_m(\beta, \gamma)$ and $\mathcal{I}_n(\beta, \gamma)$. $\mathcal{I}_m(\beta, \gamma)$ should have a change (which is the exclusive change) within the duration $[\nu, \eta]$, whereas, $\mathcal{I}_n(\beta, \gamma)$ should not contain any change-points. Consequently, we should observe a break in $p_i$ within the duration $[\nu, \eta]$ (when the auxiliary micro-activity $a_m$ was performed), but there will be no break in $p_j$. Thus, these two cases are easily separable based on the exclusive changes at $\mathcal{U}_m$; therefore, \ourmethod{} maps $p_i$ to $\mathcal{U}_m$ and $p_j$ to $\mathcal{U}_n$ unanimously for the window $[\beta,\gamma]$.

The above approach works well unless $p_i = p_j$, i.e., both $\mathcal{U}_m$ and $\mathcal{U}_n$ performs the same activity at the same instance of time. However, it is typically a rare event when multiple users within a smart environment perform the same activity simultaneously. Therefore, we argue that \ourmethod{} can label the IMU data for the majority of the cases, which is also evident from the detailed experiments, as discussed next.

%Simply put, the sole purpose of extracting these segments with \textit{exclusive changes} is to identify those windows of opportunity within the entire duration of activities where one user, say $\mathcal{U}_m$ has stopped and the other user, say $\mathcal{U}_n$, is still performing the activity. Once we obtain a \textit{exclusive change} for a user, we then identify the activity performed in the entire interval $[\alpha,\beta]$, using an audio-based activity recognition framework. 

\subsection{Putting It All Together}
We repeat the two steps mentioned above for the entire duration to obtain the activity mappings for all the \textit{exclusive changes} across both the users. Specifically, we first create a list of \textit{exclusive changes} $\mathcal{E}_m$ and $\mathcal{E}_n$ for the users $\mathcal{U}_m$ and $\mathcal{U}_n$, respectively. Mathematically, the contents of these lists can be defined as follows.
\begin{equation}
\mathcal{E}_j = \{[\theta,\zeta]| \forall\: [\nu,\eta] \in \textrm{exclusive changes in}\: \mathcal{I}_j\},\: j = \{m,n\}
\end{equation}
Once these lists are obtained for each user, we then separately get the activity mappings for each of the entries in $\mathcal{E}_m$ and $\mathcal{E}_n$ by querying the audio-based activity recognition model~\footnote{We choose the activity label with maximum confidence.} with the audio segment corresponding to the given time-interval $[\theta,\zeta]$. However, it can be noted that due to the noise in the acoustic data as well as the confusion with the acoustic-based activity recognition model, different exclusive changes from $\mathcal{E}_m$ may result in different activity labels for the other user $\mathcal{U}_n$. However, we consider that a user performs a single primary activity within the entire duration $[0, \mathcal{T}]$. Therefore, to map a single primary activity label to the user $\mathcal{U}_n$, we consider the activity label with the majority from to the list of \textit{exclusive changes} $\mathcal{E}_m$ for the user $\mathcal{U}_m$. 

\subsection{Conflict Resolution}
Although this mapping seems straightforward, several challenges may appear while mapping the activity labels to the users. Notably, we observe that a critical condition may arise when there is a conflict, and the same activity gets mapped to both the users. This appears when both the users have rarely performed the auxiliary micro-activities, or the acoustic context changes are because of the users' external noise or behavioral changes. Since \ourmethod{} is concerned with generating annotations for the IMU data for which we use a pre-trained acoustic model, fine-tuning the audio-based activity recognition model is entirely out of scope. However, where the results are conflicting, we apply the following strategy. For resolving the conflict where \ourmethod{} maps the same activity to both the users, we start by defining the \textit{opportunistic user} from the set of two users in the environment. 
\begin{definition}[Opportunistic User]\label{def:opportunistic_user}
	Among two users $\mathcal{U}_m$ and $\mathcal{U}_n$ performing simultaneous activities, $\mathcal{U}_m$ is called to be an \textit{opportunistic user} if $|\mathcal{E}_m| > |\mathcal{E}_n|$, i.e., $\mathcal{I}_m(0, \mathcal{T})$ indicates higher number of \textit{exclusive changes} than $\mathcal{I}_n(0, \mathcal{T})$.
\end{definition}
\noindent In other words, an \textit{opportunistic user} is the user who has taken more number of breaks during her primary activity and thus provided the framework more opportunities to map the activities to the other user correctly. Subsequently, to resolve the conflict, we assume the decision made by observing the changes in $\mathcal{E}_m$ to be true and map the inferred activity to $\mathcal{U}_n$.

\section{Stage 3: IMU Annotation}
\label{sec:anno}
The output of Stage 2 of the framework is the individual activity labels (corresponding to their primary activities) for each user for the entire duration $[0, \mathcal{T}]$. Next, the final task is to annotate the unlabeled IMU data for both the users with the respective mapped activity labels. It can be noted that $\mathcal{I}_m(0, \mathcal{T})$ and $\mathcal{I}_n(0, \mathcal{T})$ may contain the instances of the auxiliary activities performed by the users; however, we are only interested in annotating the IMU segments when they have performed their primary activities. Although we have a unique activity to user mapping for both users, we cannot use this mapping directly for the IMU annotation because of the following two reasons. (1) The activity labels are returned by a pre-trained acoustic-based activity recognition model~\cite{ubicoustics}, and thus the returned activity labels are synchronized with the acoustic change-points. (2) The IMU change-points may not be perfectly time-synchronized with the acoustic change-points, as they are computed independently. 

\ourmethod{} solves these issues as follows. The pre-trained acoustic-based HAR model~\cite{ubicoustics} returns the activity labels with an associated confidence value. We first find out the acoustic segment, say $\bar{\mathcal{A}}(\beta, \gamma)$, which returns the primary activity $p_i \rightarrow \mathcal{U}_m$ with the maximum confidence label. We then extract the corresponding IMU segment $\mathcal{I}_m(\beta, \gamma)$, termed as the \textit{Key Segment}. We first label the key segment $\mathcal{I}_m(\beta, \gamma)$ with the activity label $p_i$. 
%\begin{figure}[!ht]
%	\centering
%	\includegraphics[width=0.80\columnwidth,keepaspectratio]{figures/nn_plot.png}
%	\caption{Nearest neighbor approach used to annotate the user data using a known pattern as the test point}
%	\label{fig:nn_plot}
%\end{figure}
Based on the IMU change-point detection technique, we have segmented $\mathcal{I}_m(0, \mathcal{T})$ where IMU change-points mark the start and the end of each segment. We first fit all these IMU segments $\mathcal{I}_m(\theta, \zeta)$ to an unsupervised nearest neighbor approach~\cite{Kramer2013}, which learns the patterns of these IMU segments. As similar types of segments should come closer, and they collectively indicate the same activity (because the same activity should produce a similar variety of signal patterns for a user). We now use the key segment $\mathcal{I}_m(\beta, \gamma)$ to find out the $z$ number of nearest neighbors of that segment and annotate all those segments using the activity label $p_i$. The following section empirically analyzes the impact of different $z$ values.

%We achieve this by identifying the activity change patterns for the IMU data from the audio window representing that activity with the highest confidence and use to make further predictions. For example, say for an audio window $[\theta,\zeta]$ the activity label 'chopping' is predicted with the highest confidence. We then use the IMU pattern for this entire time window and use it to identify similar patterns for chopping in the IMU data, windowed by changes, using the standard unsupervised nearest neighbor approach~\cite{Kramer2013} (See \figurename~\ref{fig:nn_plot}). We then choose the closest $k$ neighbors and label all the IMU data pertaining to that window with the same activity label (say Chopping in this case).
\section{Evaluation}
\label{evaluation}
\begin{table}[]
	\scriptsize
	\centering
	\caption{Activity to User Mapping -- Workshop. Rows highlighted in red depict the erroneous cases.}
	\label{table:evaluation_main_workshop}
	\begin{tabular}{|c|c|c|l|c|}
		\hline
		&
		\multicolumn{2}{c|}{\textbf{Ground-Truth Activity Mappings}} &
		\multicolumn{2}{c|}{} \\ \cline{2-3}
		\multirow{-2}{*}{\textbf{Id}} & \textbf{Hammer} & \textbf{Saw} & \multicolumn{2}{c|}{\multirow{-2}{*}{\textbf{Output Mapping}}} \\ \hline
		C1 &
		U1 &
		U3 &
		U1: Hammer &
		\multicolumn{1}{l|}{U3: Saw} \\ \hline
		C2 &
		U1 &
		U2 &
		U1: Hammer &
		-- \\ \hline
		\rowcolor[HTML]{E06666} 
		C3 &
		U1 &
		U4 &
		U1: Saw &
		\multicolumn{1}{l|}{\cellcolor[HTML]{E06666}U4: Hammering} \\ \hline
		C4 &
		U3 &
		U1 &
		U1: Saw &
		-- \\ \hline
		C5 &
		U4 &
		U1 &
		\multicolumn{1}{c|}{--} &
		\multicolumn{1}{l|}{U4: Hammer} \\ \hline
		C6 &
		U2 &
		U1 &
		U2: Hammer &
		-- \\ \hline
		C7 &
		U4 &
		U3 &
		U4: Hammer &
		-- \\ \hline
		\rowcolor[HTML]{E06666} 
		C8 &
		U4 &
		U2 &
		\multicolumn{1}{c|}{\cellcolor[HTML]{E06666}--} &
		-- \\ \hline
		C9 &
		U2 &
		U3 &
		U3: Saw &
		-- \\ \hline
		C10 &
		U3 &
		U2 &
		U3: Hammer &
		-- \\ \hline
		C11 &
		U3 &
		U4 &
		U4: Saw &
		-- \\ \hline
		C12 &
		U2 &
		U4 &
		U2: Hammer &
		-- \\ \hline
	\end{tabular}
\end{table}

\begin{table}[]
	\centering
	\scriptsize
	\caption{Activity to User Mapping -- Kitchen. Rows highlighted in red depict the erroneous cases. Cho -- Chopping and Coo -- Cooking}
	\label{table:evaluation_main_kitchen}
	\begin{tabular}{|c|c|c|c|c|c|c|}
		\hline
		&
		\multicolumn{2}{c|}{\textbf{\begin{tabular}[c]{@{}c@{}}Ground-Truth\\ Activity\\ Mappings\end{tabular}}} &
		\multicolumn{2}{c|}{} &
		&
		\\ \cline{2-3}
		\multirow{-2}{*}{\textbf{Id}} &
		\textbf{Cooking} &
		\textbf{Chopping} &
		\multicolumn{2}{c|}{\multirow{-2}{*}{\textbf{\begin{tabular}[c]{@{}c@{}}Output\\ Mapping\end{tabular}}}} &
		\multirow{-2}{*}{\textbf{\begin{tabular}[c]{@{}c@{}}Opportunistic\\ User\end{tabular}}} &
		\multirow{-2}{*}{\textbf{\begin{tabular}[c]{@{}c@{}}Revised\\ Mapping\end{tabular}}} \\ \hline
		C1 &
		U3 &
		U1 &
		\begin{tabular}[c]{@{}c@{}}U1:\\ Cho\end{tabular} &
		\begin{tabular}[c]{@{}c@{}}U3:\\ Coo\end{tabular} &
		&
		\\ \hline
		\rowcolor[HTML]{E06666} 
		C2 &
		U2 &
		U1 &
		\begin{tabular}[c]{@{}c@{}}U2:\\ Coo\end{tabular} &
		\begin{tabular}[c]{@{}c@{}}U1:\\ Coo\end{tabular} &
		U2 &
		\begin{tabular}[c]{@{}c@{}}U1:\\ Coo\end{tabular} \\ \hline
		C3 &
		U3 &
		U2 &
		\begin{tabular}[c]{@{}c@{}}U3:\\ Coo\end{tabular} &
		\begin{tabular}[c]{@{}c@{}}U2:\\ Coo\end{tabular} &
		U2 &
		\begin{tabular}[c]{@{}c@{}}U3:\\ Coo\end{tabular} \\ \hline
		\rowcolor[HTML]{E06666} 
		C4 &
		U4 &
		U2 &
		-- &
		-- &
		\multicolumn{1}{l|}{\cellcolor[HTML]{E06666}} &
		\multicolumn{1}{l|}{\cellcolor[HTML]{E06666}} \\ \hline
		C5 &
		U4 &
		U1 &
		\begin{tabular}[c]{@{}c@{}}U4:\\ Coo\end{tabular} &
		-- &
		&
		\\ \hline
	\end{tabular}
\end{table}
We evaluate the performance of \ourmethod{} in Kitchen and Workshop scenario (see Section~\ref{datasets} for dataset) from multiple perspectives, starting with the quality of activity labels generated by \ourmethod{}, to its utility in developing supervised models, with a glimpse on its resource consumption.
\subsection{Performance of Activity to User Mapping}
%One of the primary tasks of \ourmethod{} is first to identify which user is doing what without the involvement of any human or any visual cue. 
\tablename~\ref{table:evaluation_main_workshop} and \tablename~\ref{table:evaluation_main_kitchen} indicate that \ourmethod{} can correctly map activities to users for $10$ out of $12$ cases in the Workshop environment and $3$ out of $5$ cases in the Kitchen environment, respectively. Additionally, we observe that the conflict resolution technique used in \ourmethod{} helps resolve Case C3 in the Kitchen environment.

Evidently, \ourmethod{} assigns incorrect activities to the users for combination C3 and C8 in \textbf{Workshop} and C2 and C4 in the Kitchen environment. Close inspection reveals that in $C3$ for \textbf{Workshop}, the user $U4$ performed sawing at a stretch, taking only a single break in the entire period.  This provides fewer opportunities to \ourmethod{} for annotating user $U1$ (performing hammering). On the other hand, in $C8$, the framework cannot provide any conclusive activity annotation for both the users, mostly due to the external noises present in the environment. Albeit \ourmethod{} could identify $41$ acoustic changes, the audio-based activity recognition model fails to recognize any meaningful activities for $38$ of them, resulting in a drop in performance. Side by side, we notice a general performance drop of \ourmethod{} in \textbf{Kitchen} environment. This attributes to the fact that the kitchen, in general, exhibits an inherently complex nature of activities, such as cooking and chopping. This results in a jumbled patterns in the acoustic contexts, which adversely affects the audio-based activity recognition model of the framework. For example, we observe that the model incorrectly detects cooking using the sound of handling utensils, which can occasionally occur while chopping as well.
\begin{table}
	\centering
	\scriptsize
	\caption{Accuracy and Volume of Annotations -- Workshop}
	\label{table:workshop_acc_vol}
	\begin{tabular}{|c|c|c|c|c|c|c|c|c|c|c|}
		\hline
		\multirow{3}{*}{\textbf{Id}} &
		\multicolumn{8}{c|}{\textbf{\% Annotation Accuracy {[}\% Annotation Volume{]}}} \\ \cline{2-9} 
		&
		\multicolumn{2}{c|}{\textbf{$\mathbf{z}$ = 9}} &
		\multicolumn{2}{c|}{\textbf{$\mathbf{z}$ = 11}} &
		\multicolumn{2}{c|}{\textbf{$\mathbf{z}$ = 13}} &
		\multicolumn{2}{c|}{\textbf{$\mathbf{z}$ = 15}} \\ \cline{2-9} 
		&
		\textbf{Ham.} &
		\textbf{Saw} &
		\textbf{Ham.} &
		\textbf{Saw} &
		\textbf{Ham.} &
		\textbf{Saw} &
		\textbf{Ham.} &
		\textbf{Saw} \\ \hline
		C1 &
		\begin{tabular}[c]{@{}c@{}}59.2\\ {[}20.5{]}\end{tabular} &
		\begin{tabular}[c]{@{}c@{}}95.6\\ {[}10.5{]}\end{tabular} &
		\begin{tabular}[c]{@{}c@{}}62.7\\ {[}22.8{]}\end{tabular} &
		\begin{tabular}[c]{@{}c@{}}96.3\\ {[}12.4{]}\end{tabular} &
		\begin{tabular}[c]{@{}c@{}}59.6\\ {[}25.1{]}\end{tabular} &
		\begin{tabular}[c]{@{}c@{}}96.9\\ {[}15.1{]}\end{tabular} &
		\begin{tabular}[c]{@{}c@{}}57.8\\ {[}25.9{]}\end{tabular} &
		\begin{tabular}[c]{@{}c@{}}94.7\\ {[}15.8{]}\end{tabular} \\ \hline
		C2 &
		\begin{tabular}[c]{@{}c@{}}67.9\\ {[}34.6{]}\end{tabular} &
		\begin{tabular}[c]{@{}c@{}}90.9\\ {[}29.8{]}\end{tabular} &
		\begin{tabular}[c]{@{}c@{}}70.5\\ {[}44.2{]}\end{tabular} &
		\begin{tabular}[c]{@{}c@{}}88.4\\ {[}31.8{]}\end{tabular} &
		\begin{tabular}[c]{@{}c@{}}67.3\\ {[}66.1{]}\end{tabular} &
		\begin{tabular}[c]{@{}c@{}}79.9\\ {[}45.2{]}\end{tabular} &
		\begin{tabular}[c]{@{}c@{}}61.3\\ {[}81.4{]}\end{tabular} &
		\begin{tabular}[c]{@{}c@{}}84.3\\ {[}57.6{]}\end{tabular} \\ \hline
		C4 &
		\begin{tabular}[c]{@{}c@{}}100.0\\ {[}17.0{]}\end{tabular} &
		\begin{tabular}[c]{@{}c@{}}80.4\\ {[}9.5{]}\end{tabular} &
		\begin{tabular}[c]{@{}c@{}}95.8\\ {[}17.8{]}\end{tabular} &
		\begin{tabular}[c]{@{}c@{}}80.9\\ {[}11.7{]}\end{tabular} &
		\begin{tabular}[c]{@{}c@{}}93.9\\ {[}18.5{]}\end{tabular} &
		\begin{tabular}[c]{@{}c@{}}84.7\\ {[}14.6{]}\end{tabular} &
		\begin{tabular}[c]{@{}c@{}}93.2\\ {[}36.2{]}\end{tabular} &
		\begin{tabular}[c]{@{}c@{}}76.9\\ {[}16.0{]}\end{tabular} \\ \hline
		C5 &
		\begin{tabular}[c]{@{}c@{}}98.1\\ {[}29.2{]}\end{tabular} &
		\begin{tabular}[c]{@{}c@{}}34.3\\ {[}4.4{]}\end{tabular} &
		\begin{tabular}[c]{@{}c@{}}98.2\\ {[}30.7{]}\end{tabular} &
		\begin{tabular}[c]{@{}c@{}}59.2\\ {[}11.3{]}\end{tabular} &
		\begin{tabular}[c]{@{}c@{}}97.1\\ {[}31.5{]}\end{tabular} &
		\begin{tabular}[c]{@{}c@{}}68.3\\ {[}14.6{]}\end{tabular} &
		\begin{tabular}[c]{@{}c@{}}97.5\\ {[}37.2{]}\end{tabular} &
		\begin{tabular}[c]{@{}c@{}}68.2\\ {[}15.7{]}\end{tabular} \\ \hline
		C6 &
		\begin{tabular}[c]{@{}c@{}}81.7\\ {[}4.8{]}\end{tabular} &
		\begin{tabular}[c]{@{}c@{}}87.3\\ {[}13.4{]}\end{tabular} &
		\begin{tabular}[c]{@{}c@{}}81.1\\ {[}6.6{]}\end{tabular} &
		\begin{tabular}[c]{@{}c@{}}80.1\\ {[}14.9{]}\end{tabular} &
		\begin{tabular}[c]{@{}c@{}}77.9\\ {[}7.4{]}\end{tabular} &
		\begin{tabular}[c]{@{}c@{}}81.5\\ {[}16.0{]}\end{tabular} &
		\begin{tabular}[c]{@{}c@{}}61.7\\ {[}9.9{]}\end{tabular} &
		\begin{tabular}[c]{@{}c@{}}80.1\\ {[}16.8{]}\end{tabular} \\ \hline
		C7 &
		\begin{tabular}[c]{@{}c@{}}98.7\\ {[}22.1{]}\end{tabular} &
		\begin{tabular}[c]{@{}c@{}}100.0\\ {[}26.5{]}\end{tabular} &
		\begin{tabular}[c]{@{}c@{}}97.9\\ {[}31.7{]}\end{tabular} &
		\begin{tabular}[c]{@{}c@{}}100.0\\ {[}28.8{]}\end{tabular} &
		\begin{tabular}[c]{@{}c@{}}97.9\\ {[}42.9{]}\end{tabular} &
		\begin{tabular}[c]{@{}c@{}}99.9\\ {[}29.5{]}\end{tabular} &
		\begin{tabular}[c]{@{}c@{}}97.1\\ {[}43.7{]}\end{tabular} &
		\begin{tabular}[c]{@{}c@{}}99.9\\ {[}34.0{]}\end{tabular} \\ \hline
		C9 &
		\begin{tabular}[c]{@{}c@{}}56.4\\ {[}14.9{]}\end{tabular} &
		\begin{tabular}[c]{@{}c@{}}90.6\\ {[}9.1{]}\end{tabular} &
		\begin{tabular}[c]{@{}c@{}}56.1\\ {[}15.8{]}\end{tabular} &
		\begin{tabular}[c]{@{}c@{}}91.9\\ {[}9.8{]}\end{tabular} &
		\begin{tabular}[c]{@{}c@{}}57.7\\ {[}17.3{]}\end{tabular} &
		\begin{tabular}[c]{@{}c@{}}94.5\\ {[}15.4{]}\end{tabular} &
		\begin{tabular}[c]{@{}c@{}}59.5\\ {[}18.1{]}\end{tabular} &
		\begin{tabular}[c]{@{}c@{}}94.9\\ {[}16.6{]}\end{tabular} \\ \hline
		C10 &
		\begin{tabular}[c]{@{}c@{}}92.9\\ {[}59.3{]}\end{tabular} &
		\begin{tabular}[c]{@{}c@{}}95.9\\ {[}60.3{]}\end{tabular} &
		\begin{tabular}[c]{@{}c@{}}92.2\\ {[}66.9{]}\end{tabular} &
		\begin{tabular}[c]{@{}c@{}}94.5\\ {[}62.2{]}\end{tabular} &
		\begin{tabular}[c]{@{}c@{}}88.3\\ {[}74.7{]}\end{tabular} &
		\begin{tabular}[c]{@{}c@{}}93.1\\ {[}64.2{]}\end{tabular} &
		\begin{tabular}[c]{@{}c@{}}87.3\\ {[}76.7{]}\end{tabular} &
		\begin{tabular}[c]{@{}c@{}}89.1\\ {[}67.1{]}\end{tabular} \\ \hline
		C11 &
		\begin{tabular}[c]{@{}c@{}}83.8\\ {[}33.4{]}\end{tabular} &
		\begin{tabular}[c]{@{}c@{}}100.0\\ {[}6.7{]}\end{tabular} &
		\begin{tabular}[c]{@{}c@{}}83.2\\ {[}34.1{]}\end{tabular} &
		\begin{tabular}[c]{@{}c@{}}100.0\\ {[}9.6{]}\end{tabular} &
		\begin{tabular}[c]{@{}c@{}}85.3\\ {[}39.0{]}\end{tabular} &
		\begin{tabular}[c]{@{}c@{}}100.0\\ {[}11.6{]}\end{tabular} &
		\begin{tabular}[c]{@{}c@{}}85.6\\ {[}40.1{]}\end{tabular} &
		\begin{tabular}[c]{@{}c@{}}100.0\\ {[}13.3{]}\end{tabular} \\ \hline
		C12 &
		\begin{tabular}[c]{@{}c@{}}38.6\\ {[}11.6{]}\end{tabular} &
		\begin{tabular}[c]{@{}c@{}}100.0\\ {[}6.3{]}\end{tabular} &
		\begin{tabular}[c]{@{}c@{}}38.2\\ {[}12.6{]}\end{tabular} &
		\begin{tabular}[c]{@{}c@{}}100.0\\ {[}10.9{]}\end{tabular} &
		\begin{tabular}[c]{@{}c@{}}36.2\\ {[}13.2{]}\end{tabular} &
		\begin{tabular}[c]{@{}c@{}}100.0\\ {[}12.6{]}\end{tabular} &
		\begin{tabular}[c]{@{}c@{}}38.2\\ {[}14.2{]}\end{tabular} &
		\begin{tabular}[c]{@{}c@{}}100.0\\ {[}16.2{]}\end{tabular} \\ \hline
	\end{tabular}
\end{table}

\begin{table}
			\centering
	\scriptsize
	\caption{Accuracy and Volume of Annotations -- Kitchen}
	\label{table:kitchen_acc_vol}
	\begin{tabular}{|c|c|c|c|c|c|c|c|c|}
		\hline
		\multirow{3}{*}{\textbf{Id}} &
		\multicolumn{8}{c|}{\textbf{\% Annotation Accuracy {[}\% Annotation Volume{]}}} \\ \cline{2-9} 
		&
		\multicolumn{2}{c|}{\textbf{z = 9}} &
		\multicolumn{2}{c|}{\textbf{z = 11}} &
		\multicolumn{2}{c|}{\textbf{z = 13}} &
		\multicolumn{2}{c|}{\textbf{z = 15}} \\ \cline{2-9} 
		&
		\textbf{Cook} &
		\textbf{Chop} &
		\textbf{Cook} &
		\textbf{Chop} &
		\textbf{Cook} &
		\textbf{Chop} &
		\textbf{Cook} &
		\textbf{Chop} \\ \hline
		C1 &
		\begin{tabular}[c]{@{}c@{}}100.0\\ {[}3.6{]}\end{tabular} &
		\begin{tabular}[c]{@{}c@{}}100.0\\ {[}4.8{]}\end{tabular} &
		\begin{tabular}[c]{@{}c@{}}100.0\\ {[}5.2{]}\end{tabular} &
		\begin{tabular}[c]{@{}c@{}}100.0\\ {[}6.6{]}\end{tabular} &
		\begin{tabular}[c]{@{}c@{}}86.6\\ {[}6.3{]}\end{tabular} &
		\begin{tabular}[c]{@{}c@{}}100.0\\ {[}7.7{]}\end{tabular} &
		\begin{tabular}[c]{@{}c@{}}87.3\\ {[}6.6{]}\end{tabular} &
		\begin{tabular}[c]{@{}c@{}}100.0\\ {[}8.1{]}\end{tabular} \\ \hline
		C3 &
		\begin{tabular}[c]{@{}c@{}}100.0\\ {[}3.6{]}\end{tabular} &
		\begin{tabular}[c]{@{}c@{}}100.0\\ {[}8.9{]}\end{tabular} &
		\begin{tabular}[c]{@{}c@{}}100.0\\ {[}5.1{]}\end{tabular} &
		\begin{tabular}[c]{@{}c@{}}100.0\\ {[}10.1{]}\end{tabular} &
		\begin{tabular}[c]{@{}c@{}}100.0\\ {[}6.3{]}\end{tabular} &
		\begin{tabular}[c]{@{}c@{}}100.0\\ {[}12.1{]}\end{tabular} &
		\begin{tabular}[c]{@{}c@{}}100.0\\ {[}6.6{]}\end{tabular} &
		\begin{tabular}[c]{@{}c@{}}100.0\\ {[}13.2{]}\end{tabular} \\ \hline
		C5 &
		\begin{tabular}[c]{@{}c@{}}100.0\\ {[}5.7{]}\end{tabular} &
		\begin{tabular}[c]{@{}c@{}}98.6\\ {[}9.6{]}\end{tabular} &
		\begin{tabular}[c]{@{}c@{}}100.0\\ {[}6.1{]}\end{tabular} &
		\begin{tabular}[c]{@{}c@{}}99.0\\ {[}14.3{]}\end{tabular} &
		\begin{tabular}[c]{@{}c@{}}99.9\\ {[}7.4{]}\end{tabular} &
		\begin{tabular}[c]{@{}c@{}}96.7\\ {[}16.7{]}\end{tabular} &
		\begin{tabular}[c]{@{}c@{}}94.5\\ {[}8.1{]}\end{tabular} &
		\begin{tabular}[c]{@{}c@{}}97.1\\ {[}19.1{]}\end{tabular} \\ \hline
	\end{tabular}
\end{table}
\subsection{Performance of Annotating IMU Data}
Next, we delve deep and investigate the performance of \ourmethod{} in correctly annotating the IMU stream $\mathcal{I}_m$, generated by a specific user $\mathcal{U}_m$, with a correct activity label $p_i$ (say, $p_i \rightarrow \mathcal{U}_m$). We measure the accuracy of the annotations for a user $\mathcal{U}_m$  by comparing the overlap of the annotated instances with the ground-truth activity labels for each time instance. For every overlap, we assign a score of $1$ and a $0$, otherwise.

Concerning the Workshop dataset, from \tablename~\ref{table:workshop_acc_vol}, we observe that for most of the users, the framework generates annotations (for some value of $z$) with accuracy $>70\%$, albeit the volume of annotations is less than $20\%$ for some instances, especially in the case of users performing sawing. The IMU data from sawing exhibits frequent change-points, and the IMU change patterns within those change-points depend on factors like the holding the saw and the speed. Consequently, we see significant variations in the per-window IMU pattern. As we rely on a nearest-neighbor strategy for labeling the IMU data, such variations in the patterns result in few IMU windows showing similarity with the key segment (Section~\ref{sec:anno}), resulting in a lower volume of annotated data. 

However, for Kitchen (see \tablename~\ref{table:kitchen_acc_vol}), we observe that \ourmethod{} performs with better accuracy. This improvement can be attributed to the inherent nature of kitchen activities like cooking and chopping, which by default require intermediate stops, thus providing many opportunities for the framework to identify the patterns accurately. However, one important point regarding the analysis and similarity of patterns also reveals that common activities like cooking often have high variability~\cite{buildsys20}. For example, users may change the way they use the cooking spud depending on the item being cooked. These variations over time cause frequent small change windows impacting the volume of annotation to some extent. 

One critical observation is that the IMU patterns heavily depend on the activity type. For example, auxiliary micro-activities are much less when a user cuts a metal pipe. In general, such opportunities are less in number for workshop activities. On the contrary, kitchen activities provide a better opportunity for annotation. Therefore, \ourmethod{} can widely be used for annotating \textit{Activities of Daily Livings} (ADLs); nevertheless, it can also annotate non-ADL activities (like Workshop) to a satisfactory extent. 
%\begin{figure*}[]
%	\captionsetup[subfigure]{}
%	\begin{center}
%		\subfloat[\label{fig:hammer_dist}]{
%			\includegraphics[width=0.25\linewidth,keepaspectratio]{figures/hammer_distances.png}
%		}
%		\subfloat[\label{fig:saw_dist}]{
%			\includegraphics[width=0.25\linewidth,keepaspectratio]{figures/saw_distances.png}
%		}
%		\subfloat[\label{fig:cooking_dist}]{
%			\includegraphics[width=0.25\linewidth,keepaspectratio]{figures/Cooking_distances.png}
%		}
%		\subfloat[\label{fig:chopping_dist}]{
%			\includegraphics[width=0.25\linewidth,keepaspectratio]{figures/Chopping_distances.png}
%		}
%	\end{center}
%	\caption{Variation of Euclidean distance across different $z$ for (a) Hammering, (b) Using saw, (c) Cooking, and (d) Chopping.}
%	\label{fig:nn_analysis}
%\end{figure*}
%
%Furthermore, an additional investigation in this direction reveals that indeed for certain activities like using saw and cooking the patterns change over a time (see \figurename~\ref{fig:nn_analysis}) which leads to larger Euclidean distances for the set of unsupervised nearest neighbors, albeit without hampering the overall accuracy however, with a decrease in volume as explained above.
\subsection{Benchmarking Annotated Data}
\begin{figure}[]
	\captionsetup[subfigure]{}
	\begin{center}
		\subfloat[\label{fig:workshop_cross_user}]{
			\includegraphics[width=0.59\linewidth,keepaspectratio]{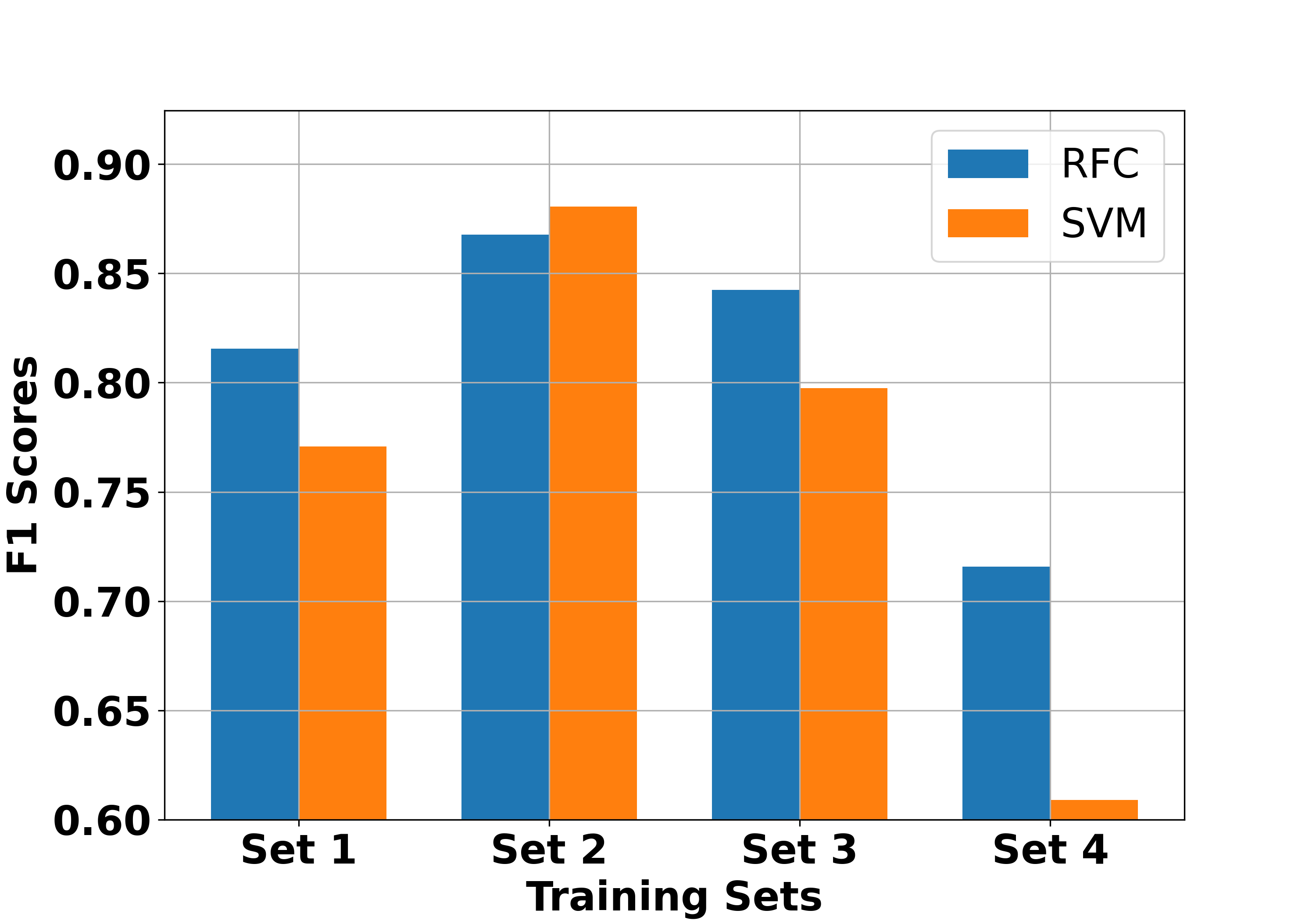}
		}
		\subfloat[\label{fig:kitchen_cross_user}]{
			\includegraphics[width=0.37\linewidth,keepaspectratio]{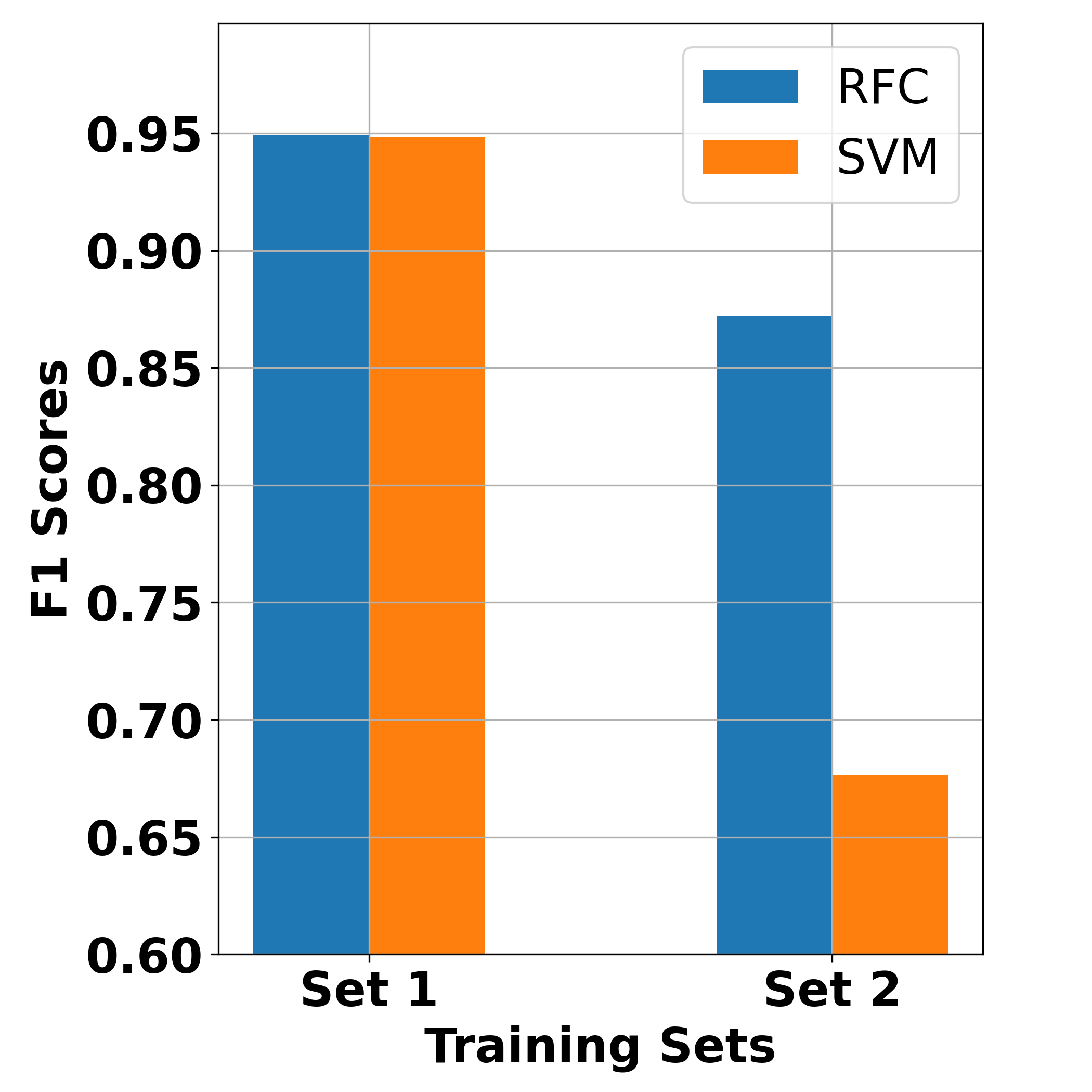}
		}
	\end{center}
	\caption{Cross-user activity recognition accuracy with the annotated data produced by the framework in the (a) Workshop and (b) Kitchen setups, respectively.}
	\label{fig:benchmark}
\end{figure}

The primary purpose of annotating sensor data is to create training data for supervised models. Understanding this objective, we assess the quality of our annotated data in a cross-user setup. This setup allows us to benchmark the annotated data and provides insight if \ourmethod{} can be used to get bootstrap data for a few users and then whether such bootstrapping data can be used to obtain the label for other unlabeled data. To evaluate this, we use a leave-one-out strategy to test a particular users' original data with supervised models trained using the annotated labels ($z = 15$) and the labeled IMU streams as features from the other users in the setup. For the Workshop, we thus generate four sets (Set 1 to Set 4) of training data by leaving out one of the four users in each of the cases and use the left-out user to test the model accuracy. For the Kitchen, we had data for both the activities only for user $U2$ whose data we use as the test dataset and create two training datasets with a mixture of chopping data from $U1$ and cooking data from users $U3$ (Set 1) and $U4$ (Set 2), respectively. From the results with two standard supervised learning algorithms (Random Forest and Support Vector Machines) shown in \figurename~\ref{fig:benchmark}, we observe that for most of the cases the supervised models attain $>0.70$ F\textsubscript{1}-score, in correctly predicting the activity labels in the cross-user setup, which demonstrates the quality of annotation by \ourmethod{}.

\subsection{Resource and Time Consumption}
\begin{figure}[]
	\captionsetup[subfigure]{}
	\begin{center}
		\subfloat[\label{fig:memory_consumption}]{
			\includegraphics[width=0.50\linewidth,keepaspectratio]{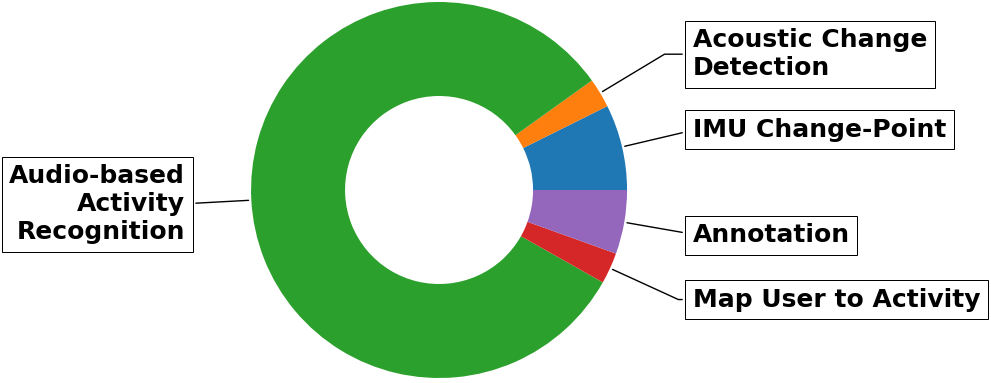}
		}
		\subfloat[\label{fig:time_consumption}]{
			\includegraphics[width=0.48\linewidth,keepaspectratio]{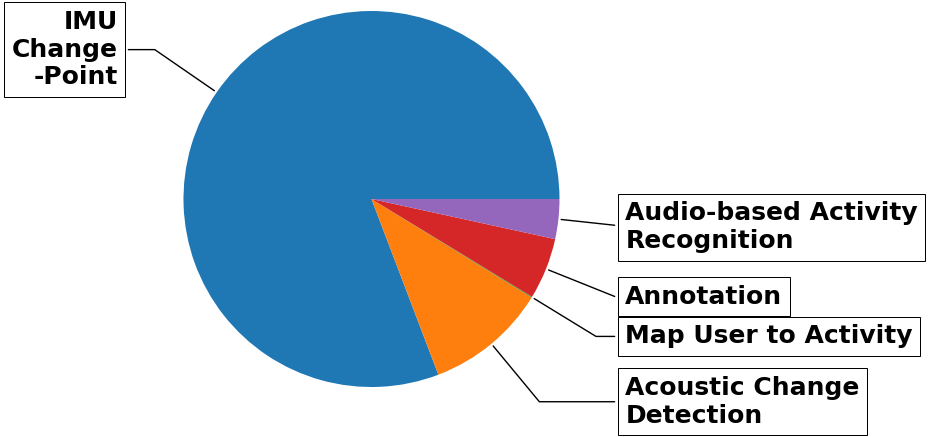}
		}
	\end{center}
	\caption{Framework -- (a) Memory and (b) Time profiling.}
\end{figure}

\ourmethod{} uses audio as an auxiliary modality to make the overall framework lightweight so that the privacy-sensitive data can be kept within its territory by running the model on the edge. To assess this, we profile the stages of the framework individually on a per-module basis for time and memory consumption (using Linux \texttt{proc} filesystem) for the C1 dataset from the Workshop setup. \figurename~\ref{fig:memory_consumption} shows that except for the audio activity-recognition module, all the remaining modules take $<200$MB memory which allows them to run on resource-constrained edge devices. Albeit the audio-based HAR module engrosses a significant volume of memory, however, it can run on resource-constrained devices as shown in~\cite{ubicoustics}. Concerning the time consumption, the total running time for \ourmethod{} on Workshop-C1 is $969.15$seconds, with the majority of time being consumed to detect changes in the IMU data. However, this total time consumed is for the entire dataset containing the IMU streams from both users, which may be provided in sessions to reduce data volume and processing time.
\section{Discussion}
This section highlights a few of the insights we got while developing \ourmethod{}, opening up various exciting issues.  

\textbf{Scalability}: We primarily designed and tested \ourmethod{} for dual user simultaneous activities. However, our method can be extended for multiple users by applying a leave-one-out policy as follows. Let there be a set of $n$ users $\mathbb{U}=\{\mathcal{U}_1, \hdots, \mathcal{U}_n\}$ and $n$ different activities $\mathbb{P} = \{p_1, \hdots, p_n\}$. Let there be an acoustic gap for $\mathcal{U}_1$, indicating that the primary activity for $\mathcal{U}_1$, say $p_1$, will be absent in that duration. So, we can recursively probe \ourmethod{} with $\mathbb{U} \setminus \{\mathcal{U}_1\}$ and $\mathbb{P} \setminus p_1$. Nevertheless, such a method needs fine tuning of activity signatures, which can be explored as an immediate extension of this work. 
%However, in real life, it is implausible that a lot of users perform different activities under a common acoustic context; therefore, such a complex variant of \ourmethod{} is hardly required in reality.  

\textbf{Change in Intra-Activity Patterns}: An unsupervised nearest neighbor allows us to identify closely related patterns, and \ourmethod{} uses this idea to annotate using a known key segment. However, one crucial observation in this context is regarding the change of patterns over time. For example, a user may change how s(he) uses the cooking spud over time depending on how well the food is cooked. Such changes over time can increase the distance of a valid instance from the known key segment, resulting in lower annotation volume. Although \ourmethod{} cannot tackle such behavioral changes, it can provide significant bootstrapping data. We can use such bootstrapping data in approaches like \textit{Active Learning}~\cite{active_main} for further annotation with minimum human intervention or adapt approaches like~\cite{buildsys20} for tackling data variability.

\textbf{Significant Acoustic Gap}: A primary understanding that we develop from the overall evaluation is that the acoustic gaps are critical for \ourmethod{} to perform well. However, a significant loud noise from such auxiliary activities may collude the overall acoustic context making it difficult to get identified by the audio-based HAR. Thus, the particular acoustic gap becomes irrelevant.

%\textbf{Impact of Augmented Data}: This paper uses an augmented dataset that mimics various diversities present, which are likely to be present in the IMU and acoustic signature captured from a real-life multi-user scenario. However, one factor that may additionally come into consideration in such a real-life multi-user scenario is the behavioral impact of one user on the other. For example, a user may give fewer or more breaks depending on how her partner is giving, which may impact the overall performance of \ourmethod. Such behavioral intricacies need further study.
\section{Conclusion}
\textit{``Data, data, everywhere, nor any drop to use.''} Smart infrastructures generate millions of sensing data per second, but the data is useless until they are appropriately labeled. Human-based annotations are not feasible and cost-effective for labeling such data, whereas video-based annotation has significant processing overhead and privacy concerns. \ourmethod{} provides a framework for automated annotation of the IMU data using audio captured through a smart speaker or a VPA deployed within the infrastructure. 
%Audio has multiple advantages. (1) Unlike a camera that captures the video from a specific direction (except 360$^{\circ}$ Cameras), a microphone has a more comprehensive range of operations; thus, deployment is not at all a concern. (2) Audio processing is much lightweight; therefore, it can be done entirely over the edge devices, thus keeping the data within its boundary. 
\ourmethod{} judicially combines signal processing techniques with unsupervised learning mechanisms to correctly identify the activity boundaries to label the IMU data with high accuracy. The evaluation indicates that \ourmethod{} generated labels are highly accurate, as the method identifies the activity boundaries by extracting the precise changes within the IMU data itself.  Although the volume of \ourmethod-labeled data is low sometimes, we believe that our approach can at least generate the bootstrap labels that can then be combined with techniques like \textit{Active Learning}~\cite{active_main} to annotate the remaining data. 
%Therefore, \ourmethod{} can be integrated with existing smart infrastructures to generate a vast volume of labeled data from diverse activities, which then can be used for developing multiple interesting applications by mining various human activity patterns.
\balance
\bibliographystyle{IEEEtran}
\bibliography{reference/ref}
\end{document}